---

# Growth, Inequality and Declining Business Dynamism in a Unified Schumpeter Mark I + II Model

Patrick Mellacher[1]

[1] University of Graz, Graz Schumpeter Centre,

Universitätsstraße 15/FE, A-8010 Graz

**Abstract**

I develop a simple Schumpeterian agent-based model where the entry and exit of firms, their productivity and markup, the birth of new industries and the social structure of the population are endogenous and use it to study the causes of rising inequality and declining "business dynamism" since the 1980s. My hybrid model combines features of i) the so-called Schumpeter Mark I (centering around the entrepreneur), ii) the Mark II model (emphasizing the innovative capacities of firms), and iii) Cournot competition, with firms using OLS learning to estimate the market environment and the behavior of their competitors. A scenario which is quantitatively calibrated to US data on growth and inequality replicates a large number of stylized facts regarding the industry life-cycle, growth, inequality and all ten stylized facts on "declining business dynamism" proposed by Akcigit and Ates (AEJ:Macro, 2021). Counterfactual simulations show that antitrust policy is highly effective at combatting inequality and increasing business dynamism and growth, but is subject to a conflict of interest between workers and firm owners, as GDP and wages grow at the expense of profits. Technological factors, on the other hand, are much less effective in combatting declining business dynamism in my model.

# Growth, Inequality and Declining Business Dynamism in a Unified Schumpeter Mark I + II Model

Patrick Mellacher[1]

**Abstract**: I develop a simple Schumpeterian agent-based model where the entry and exit of firms, their productivity and markup, the birth of new industries and the social structure of the population are endogenous and use it to study the causes of rising inequality and declining "business dynamism" since the 1980s. My hybrid model combines features of i) the so-called Schumpeter Mark I (centering around the entrepreneur), ii) the Mark II model (emphasizing the innovative capacities of firms), and iii) Cournot competition, with firms using OLS learning to estimate the market environment and the behavior of their competitors. A scenario which is quantitatively calibrated to US data on growth and inequality replicates a large number of stylized facts regarding the industry life-cycle, growth, inequality and all ten stylized facts on "declining business dynamism" proposed by Akcigit and Ates (AEJ:Macro, 2021). Counterfactual simulations show that antitrust policy is highly effective at combatting inequality and increasing business dynamism and growth, but is subject to a conflict of interest between workers and firm owners, as GDP and wages grow at the expense of profits. Technological factors, on the other hand, are much less effective in combatting declining business dynamism in my model.

## 1 Introduction

Throughout the history of capitalism, mankind witnessed remarkable growth in productivity and product variety. This growth, however, has also been connected to persistent inequality. Ever since, economists investigated the causes, consequences and interrelation between these two key features of the prevalent economic system. While Kuznets famously proposed an inverted U-curve between economic development and inequality, i.e., that inequality would eventually decline once societies grew enough, the underlying empirical trend reversed in the 1980s (see Piketty 2014). Thus, the question of distribution, once described by David Ricardo as the "principal problem in Political Economy" (Ricardo 1817, p.1), has reclaimed a prominent place in economics in recent years.

[1] Graz Schumpeter Centre, University of Graz, patrick.mellacher@uni-graz.at

A number of different explanations have been put forward to explain the emergence, presence and increase of inequality. Although a clear distinction is not always possible, explanations may be attributed to at least one of three categories:

The first type of explanation argues that inequalities arise endogenously from economic processes that are inseparably connected to capitalism. This view is held by thinkers as important and diverse as Adam Smith, Joseph Schumpeter and Karl Marx, who approached this question from differing theoretical perspectives and normative points of view. While Marx (1890) emphasized a) the inequality *between* classes, which stems from and reproduces the "exploitation" of the working class and b) a tendency towards increasing concentration *within* the capitalist class, Smith saw inequalities arising from market processes as being "natural" and "useful" in contrast to "artificial" inequalities produced by policy (see Walraevens 2021). Schumpeter, an ardent defender of capitalism's ability to encourage innovation and growth, characterized capitalism as "the civilization of inequality and of the family fortune" (Schumpeter 2003[1942], p. 379). It is important to note that any serious economist – and especially those named – would recognize that capitalism also generates counteracting tendencies to at least some forms of inequality. If a certain business or production technology produces supernormal profits, others will try to enter this market or copy this technique and eventually drive down profits again. Nevertheless, empirical research by De Loecker et al. (2020) show that both market power and profits have increased since the 1980s.

A second explanation attributes inequality to innovation and technological change, with several potential causes. On the one hand, technological change may affect employment, which in turn feeds back to the distribution. The literature in the tradition of evolutionary economics has long emphasized the non-trivial effects of technological change on employment, as new technologies may enhance labor productivity and/or create new demand (see, e.g., Dosi 1984; Dosi et al. 2021a; Saviotti and Pyka 2004b) and have argued to be part of "long waves" of economic development (Schumpeter 1939; Freeman et al. 1982). On the other hand, innovations may change the (composition of) the social structure of society. This is emphasized by Schumpeter, who argued in his "Theory of Economic Development" (Schumpeter 1934) that entrepreneurial activity, i.e., innovation, is the main source of social mobility in capitalist societies, as successful entrepreneurs elevate by creatively destroying the position of some incumbents: "the upper strata of society are like hotels which are indeed always full of people, but people who are forever changing. They consist of persons who are recruited from below to a much greater extent than many of us are willing to admit" (Schumpeter 1934, p. 156). From the late 1990s onwards, there has also been a strong interest in empirical research on the topic (e.g., Autor et al. 1998; Autor et al. 2003), mostly focusing on specific aspects of the relationship between technological change and inequality, and often being combined with highly stylized neoclassical/equilibrium-based models (e.g., Aghion et al. 2019). The new empirical evidence has triggered a large wave of theoretical research on inequality in its many shades. For instance, evolutionary models have been applied to study technological change and labor market polarization (Bordot and Lorentz 2021; Mellacher and Scheuer 2021; Fierro et al. 2022), inequality, technological change and secular stagnation (Borsato 2022), and technological change and inequality in a multi-country setting (Dosi et al. 2019).

The third type of explanations trace inequality back to social and/or political reasons. While Adam Smith argued that governmental policy creates inequality (Walraevens 2021), governmental activity is, at least in democracies, widely perceived to have reduced inequality. Recent empirical research has shown, for instance, that minimum wage increases are able to increase wages at the lower end of the wage distribution with no or only a low effect on the level of unemployment (Cengiz et al. 2019; Derenoncourt and Montialoux 2021; Dustmann et al. 2021) and that unions significantly decrease inequality (Farber et al. 2021). This is in stark contrast to the neoclassical orthodoxy still largely taught in introductory microeconomics. Furthermore, the weakening of antitrust is connected to certain features of declining business dynamism (Vaziri 2022). More generally, one can argue that governmental activity is able to create or strengthen institutions which are either conducive or detrimental to inequality (e.g., Acemoglu and Robinson 2002; Piketty and Saez 2014). Agent-based modelling is perfectly suited to study such "institutional" setups, as shown, e.g., Dosi et al. (2018b) by who contrast a "Fordist" and a "Competitive" institutional regime.

Without neglecting or downplaying the important role of other important social and political factors in determining the level of inequality, I focus in this paper on the role of i) antitrust policies, ii) technological factors and iii) accumulation tendencies inherent to capitalist market economies. Using my benchmark scenario, which replicates rising inequality and all stylized facts on "declining business dynamism" proposed by Akcigit and Ates (2021), and three counterfactual scenarios, I show that sustained strong antitrust enforcement can reverse the rise in inequality and at least seven out of ten stylized facts on declining business dynamism. In contrast, enabling firms to imitate in a much easier way can only reverse three of the stylized facts in my model, whereas increasing the growth of technological opportunities, which allows for a stronger growth in the number of industries, can only dampen the effects of declining business dynamism.

I derive my results from a novel, rather simple Schumpeterian model that combines important features of the so-called Schumpeter Mark I model and the Mark II model (as first coined by Freeman 1982). The Mark I model, outlined in "Theory of Economic Development" (Schumpeter 1934), centers around the entrepreneur, who disrupts the circular flow of the economy by using "new combinations" (see Kurz 2012), i.e., by introducing new products, discovering new markets, improving the production processes etc. In his Mark II model, developed in "Capitalism, Socialism and Democracy" (2003[1942]), Schumpeter emphasizes the role of big corporations for technological progress. Pressured by competition, they must invest in Research & Development to continuously improve their products and cut their costs of production.

I formalize Schumpeter's theories with an agent-based model that combines key features of Schumpeter Mark I and Mark II, which are – in related literature – often seen as mutually exclusive regimes (see, e.g., Wersching 2010). In my unified model, entrepreneurs play the leading role in the birth of new industries. Entrepreneurs can start a new firm by either founding a new industry, if they spot a possibility to do so, or by imitating an existing firm. Firms invest in R&D to continuously improve their productivity and increase their profit rate, as well as push out their competitors. Firms engage in Cournot competition and the markup of each firm thus endogenously depends on the number of its competitors, as well as on its productivity relative to the productivity of the other firms in its industry. This distinction between the birth (captured by Mark I) and evolution of industries (captured by Mark II) is inspired by the stylized fact that individual entrepreneurs are often connected to major product innovations, while the following incremental innovations (that eventually create a dominating market position) are created by dedicated R&D staff. In the IT industry, the former stylized fact is epitomized by figures like Bill Gates, Steve Jobs, Mark Zuckerberg and Jeff Bezos who apparently have played a particularly important role during the early stages of their respective companies.

I further implement a modification to the well-known standard R&D mechanism developed by Dosi et al. (2010) that is potentially able *explain* the decrease in the diffusion of technological progress as identified by Akcigit and Ates (2021) to be a plausible factor causing the phenomenon of "declining business dynamism", by introducing a technological distance penalty to imitation: If the productivity gap between a firm and the industry leader is larger, it is harder for this firm to imitate. The same mechanism applies to entrepreneurs who want to enter a new industry by imitation. Thus, the role of the entrepreneur diminishes, and the role of the firm becomes more important once an industry becomes more mature. The basic mechanism, i.e., that the probability of knowledge spillovers between firms depends on their technological distance, is grounded in evolutionary theory (Perez and Soete 1988) and supported by empirical evidence (e.g., Bloom et al. 2013). More generally, this mechanism can also represent other mechanisms that may inhibit copying successful business models in more mature industries such as network effects. There seems to be plenty anecdotal evidence from the IT industry to support this claim: for instance, the microblogging website X (formerly known as Twitter) has been challenged by numerous competitors (e.g., by "Truth Social", Mastodon, Blue Sky etc.). While these entrants managed to copy most of the core functions of the market leader, they were (as of the writing of this article) ultimately unable to establish a presence beyond a market niche. After all, users will only use a microblogging service if *others* also use them, which is clearly path-dependent. Another example for complementarities impeding market entries are compatibility issues. A new competitor seeking to challenge Microsoft Windows would not only have to replicate code that took decades to develop, but also have to deal with the question that existing software is developed to be used with a specific operating system. While my model does not explicitly consider such mechanisms, the imitation distance penalty parameter can be viewed to be a proxy for them.

Finally, my results are based the comparison of a baseline scenario with various counterfactual scenarios, where the baseline scenario is obtained by fitting the model to empirical data using a neural network based approach to Approximate Bayesian computation (Blum and Francois 2010). This method outperformed other reference table algorithms studied by Carrella (2021) such as Random Forest regression in my cross-validation experiments. This approach helps to

In addition to the vast stream of literature on the causes of inequality and, in particular, its interrelation with economic growth and technological change as outlined above, my paper contributes to five other, more specialized, strands of the literature:

First, to the formal modelling of endogenous technological change at the sectoral level inspired by Schumpeterian theories, which have made a big impact on evolutionary and agent-based economics, starting with Nelson and Winter (1982) and followed by many others (for evolutionary models, see e.g., Silverberg et al. 1988; for agent-based models, see Dawid 2006 for an early survey of ABMs featuring technological change). Subsequently, evolutionary economists have managed to implement technological change in large-scale macroeconomic models (Dosi et al. 2006; Dawid et al. 2008; Deissenberg et al. 2008). In particular, the seminal Keynes+Schumpeter (K+S) model by Dosi et al. (2010) introduced endogenous technological change at the firm-level using a simple evolutionary process in the spirit of Nelson and Winter (1982) that has subsequently been applied to other macroeconomic agent-based models (ABMs) (e.g., Caiani et al. 2019; Terranova and Turco 2022). Other macroeconomic ABMs featuring endogenous technological change include the EURACE@Unibi model (Dawid et al. 2012), the LAGOM model (Wolf et al. 2013) and the model proposed by Lorentz et al. (2016). Apart from large-scale ABMs, more simple approaches such as the model by Dosi et al. (2017b) have also proven to be well-equipped to replicate important empirical stylized facts. My contribution to this literature is a) to adapt the R&D mechanism by Dosi et al. (2010) to account (in a different way) for the stylized fact that the technological distance between two firms influences the probability of knowledge spillovers as argued above, and b) to allow for the endogenous entry of new firms, depending on the availability of entrepreneurs who have enough funds to enter a new market and on their investment alternatives. Like the model proposed by Dosi et al. (2017b), my approach is rather simple and does not aim to model the economy in the level of detail provided by large-scale models such as the K+S model or the EURACE model. Models in the tradition of "mainstream" economics based on (general) equilibrium have started to embrace endogenous technological change with Romer (1990) and Aghion and Howitt (1992).

Second, I contribute to the literature on endogenous growth in the number of industries and products, most of which stand in a Schumpeterian tradition and some of which also feature a connection to inequality and/or labor market outcomes. This stream of literature has been pioneered by Saviotti and Pyka (2004a), who develop a model in which entrepreneurs endogenously create new sectors. Saviotti and Pyka (2004b; 2008) adapt this framework to study, among others, the impact of technological change on employment. In these models, new sectors are created once existing ones are "saturated" in the sense that they are devoid of monopoly profits. While I implicitly also include this mechanism as part of the short-sighted profit-maximizing behavior of entrepreneurs, my model also allows for (and witnesses) the (nearly) simultaneous creation of new sectors, depending on the availability of technological opportunities and entrepreneurial funds, as well as the level of concentration in the existing industries. Wersching (2010) develops a model in which Schumpeter Mark I and II represent mutually exclusive technological regimes, both of which are able to produce growth in the number of products. In contrast, my model integrates both Mark I and Mark II as different drivers of technological change into a unified model where the importance of both engines emerges endogenously from the model. Further related contributions include papers by Ciarli and Lorentz (2010), Savin and Egbetokun (2016), Vermeulen et al. (2018) and Gräbner and Hornykewycz (2022). The paper which is most closely related to the present work is by Dosi et al. (2022), who develop a model of endogenous creation of new sectors based on the K+S framework (Dosi et al. 2010) and employ it to study, *inter alia*, the creation and destruction of labor. While their model is more complex than mine in most respects, such as with regard to the modelling of the labor and product markets, I contribute by modelling industry entry as an explicit action by entrepreneurs that is based on a probability to imitate which depends on the technological maturity of an industry. In particular, however, my analysis has a different focus as I am interested in the replication of the stylized facts on the rise in inequality and "declining business dynamism" since the 1980s.

Third, I contribute to the literature on explaining inequality and labor market outcomes using models with endogenous technological change, which covers both contributions from agent-based and disequilibrium models (e.g., Bordot and Lorentz 2021; Borsato 2022; Caiani et al. 2019; Fierro et al. 2022; Carvalho and Di Guilmi 2020; Dawid and Hepp 2022; Dosi et al. 2017a; Dosi et al. 2018a; Dosi et al. 2019; Fanti 2021; Mellacher and Scheuer 2021; Terranova and Turco 2022) and general equilibrium theory (Acemoglu 1998; Acemoglu 2002; Aghion 2002; Aghion et al. 2019; Akcigit and Ates 2021; Jones and Kim 2018, Vaziri 2022). Among the general equilibrium models, my analysis is closest to the papers by Akcigit and Ates (2021) and Vaziri (2022). Akcigit and Ates (2021) present ten stylized facts about "declining business dynamism" and develop a simple model where a decrease in knowledge diffusion (i.e., imitation) can explain six out of these facts.[2] They further argue that this factor has the potential to explain the other four stylized facts. Vaziri (2022), on the other hand, provides evidence for three stylized facts connected to stronger antitrust enforcement, namely a higher level of firm entries and a higher level of productivity growth, but lower R&D investment. She then develops and calibrates a general equilibrium framework that replicates these stylized facts, as well as an increase in the labor share of GDP. Within the agent-based stream of this literature, my analysis is closest to Terranova and Turco (2022), who study how inequality and concentration are driven by (the absence of) knowledge spillovers building on fully-fledged macroeconomic

[2] Extending and using the model by Caiani et al. (2016), Reiter (2019) shows that a decrease in knowledge diffusion can explain at least 3 out of these stylized facts also in a fully-fledged macroeconomic ABM.

ABM by Assenza et al. (2015) and extending it, inter alia, by the R&D mechanism from Dosi et al. (2010). I add to this literature by developing the first model which is capable of replicating all of the ten stylized facts. Furthermore, I conduct counterfactual policy simulations showing that increased antitrust enforcement can reverse seven of the ten facts.

Fourth, this paper contributes to the literature using agent-based models to study inequality and labor market dynamics. In addition to the papers mentioned in the preceding paragraphs, this literature also contains explanations for inequality that do not refer primarily to technological change, but e.g. to social networks (Gemkow and Neugart 2011; Dawid and Gemkow 2014), labor market regimes, wage setting policies and de-unionization (Caiani et al. 2020; Ciarli et al. 2019; Dawid et al. 2021; Dosi et al. 2017a; Dosi et al. 2018a; Dosi et al. 2021b), disinformation and policy (Mellacher 2021), job mobility (Applegate and Janssen 2022) and more stylized approaches (e.g. Vallejos et al. 2018).

Fifth, this paper contributes to the literature on examining the problems related to concentration and monopolization and the impact of policies such as antitrust (Malerba et al. 2001; Dosi et al. 2010; Dosi et al. 2017c) and patents (Çevikarslan 2017; Dosi et al. 2023) using agent-based models. Çevikarslan (2017) and Dosi et al. (2023) both show that patent length and breadth have a strong impact on market concentration. Focusing on the evolution of the computer industry, Malerba et al. (2001) shows that the effectiveness of antitrust policies in constraining market dominance depends on the technological characteristics of the market. Dosi et al. (2010) and Dosi et al. (2017c), on the other hand, investigate the macroeconomic effects of antitrust policies in the K+S model. They find that antitrust policies are connected to a higher GDP growth rate, but do not study their impact in detail with regard to "declining business dynamism". Since multiple papers suggest a connection to declining business dynamism (e.g., Akcigit and Ates 2021, Akcigit and Ates 2023), and Vaziri (2022) finds a connection to some characteristics of declining business dynamism, my paper answers the call for a systematic analysis of antitrust policies on all of the ten facts mentioned by Akcigit and Ates (2021).

Naturally, this paper does not aim or claim to provide an exhaustive analysis of the interrelation between growth, technological change, market concentration, declining business dynamism and inequality. This is due to the fact that my simple model does not integrate all aspects that may play a role, some of which are implemented by other authors mentioned in the preceding paragraphs. My analysis is thus complementary to other approaches, and the method used in this paper, i.e., agent-based modelling, has indeed the potential to integrate multiple sources of inequality and declining business dynamism in a more general model at a later stage.

The rest of this paper is structured as follows. The second section describes the model. The third section explains how the model is calibrated to the historical baseline scenario and shows that it is able to reproduce important stylized facts on growth, industry life-cycle, inequality and declining business dynamism. It then moves on to counterfactual policy simulations to understand how the decline in business dynamism can be reversed within my model. The fourth section concludes.

## 2 Model

The model mainly draws on three sources of inspiration:

First, following Joseph Schumpeter's theories, I consider two sources of innovation: On the one hand the entrepreneur, who is at the center of attention of Schumpeter's *Theory of Economic Development* (Schumpeter 1934), and on the other hand established firms, as discussed in his *Capitalism, Socialism and Democracy* (Schumpeter 2003[1942]). In my model, entrepreneurs play the leading role during the birth of new industries, by a) radically innovating, i.e., founding new industries by exploiting technological opportunities, and b) by imitating young firms in order to reap some of their surplus profits. I call this source the Mark I engine, following the popular characterization of the *Theory of Economic Development* as Mark I by Christopher Freeman (1982, p. 212). Established firms, on the other hand, conduct research and development (R&D) in order to continuously improve their processes to improve their market position. I call this source the Mark II engine, following Freeman's (1982, p. 213) characterization of (some of) the ideas brought forward in *Capitalism, Socialism and Democracy*. I further assume, that imitation becomes more difficult with an increasing technological distance (this is grounded in evolutionary theory, see Perez and Soete 1988, and supported by empirical evidence, see e.g., Bloom et al. 2013). Hence, it becomes more difficult for entrepreneurs to enter existing markets after some time, and the importance of the Mark I engine diminishes once industries become mature and – vice versa – the importance of the Mark II engine grows over time, looking at a specific industry.

Second, following classical political economy, as well as an increasing fraction of evolutionary and agent-based macroeconomics (e.g., Assenza et al. 2015; Borsato 2022; Mellacher 2020; Rengs and Scholz-Wäckerle 2019), but also some general equilibrium models (e.g. Vaziri 2022), the agents in my model are divided into classes, namely two: workers and entrepreneurs (who are also capitalists). This is a small deviation from Schumpeter (1934), who argued that the capitalist and the entrepreneur are different actors. Also, for simplicity, and because it is not the focus of my analysis, I follow classical political economy in assuming that each goods market and the labor market is cleared during each time step of the simulation. It is important to note, however, that this does *not* imply general equilibrium or even partial equilibrium beyond the ultra-short run, as I assume that the agents are boundedly rational and imperfectly informed about the capabilities and strategies chosen by the other agents.

My third source of inspiration is evolutionary and Neo-Schumpeterian economics. Modelling technological change *within sectors* as a two-step stochastic process involving evolutionary selection was pioneered by Nelson and Winter (1982), and later refined and introduced as the standard agent-based macroeconomic approach by Dosi et al. (2010). I use an adapted version of this approach to model the Schumpeter Mark II engine, i.e., technological change *within* each sector. Furthermore, evolutionary economics has a strong record on modelling structural change in the number of industries (see, e.g., Saviotti and Pyka 2004a; Saviotti and Pyka 2004b; Saviotti and Pyka 2008; Dosi et al. 2022)

The model is agent-based, which means that it centers around explicitly modeled heterogeneous interacting agents whose behavior is described in this section. Figure 1 gives an overview of the model, which is described in detail in the following subsections. Firms employ workers (subsection 2.3). Each industry is located at one "technological opportunity" (subsection 2.4). The number of technological opportunities may grow over time due to advances in basic research, but following Schumpeter (see Freeman 1982) the latter are exogenous to the economic processes depicted in the model. Each firm is located at a single industry, where it conducts R&D (covering both invention and imitation) to improve its productivity and engages in (boundedly rational) Cournot competition (subsection 2.2). Entrepreneurs own firms and earn their profits. They consume a fraction of their funds and save the rest to finance new firms. Entrepreneurs can try to a) found new firms by radically innovating, i.e., trying to exploit an unused technological opportunity by founding a new firm in a new industry, or b) imitate an existing firm (subsection 2.5). Households, i.e., workers and entrepreneurs, consume goods from industries by splitting their consumption budget evenly among industries with a positive output (2.1).

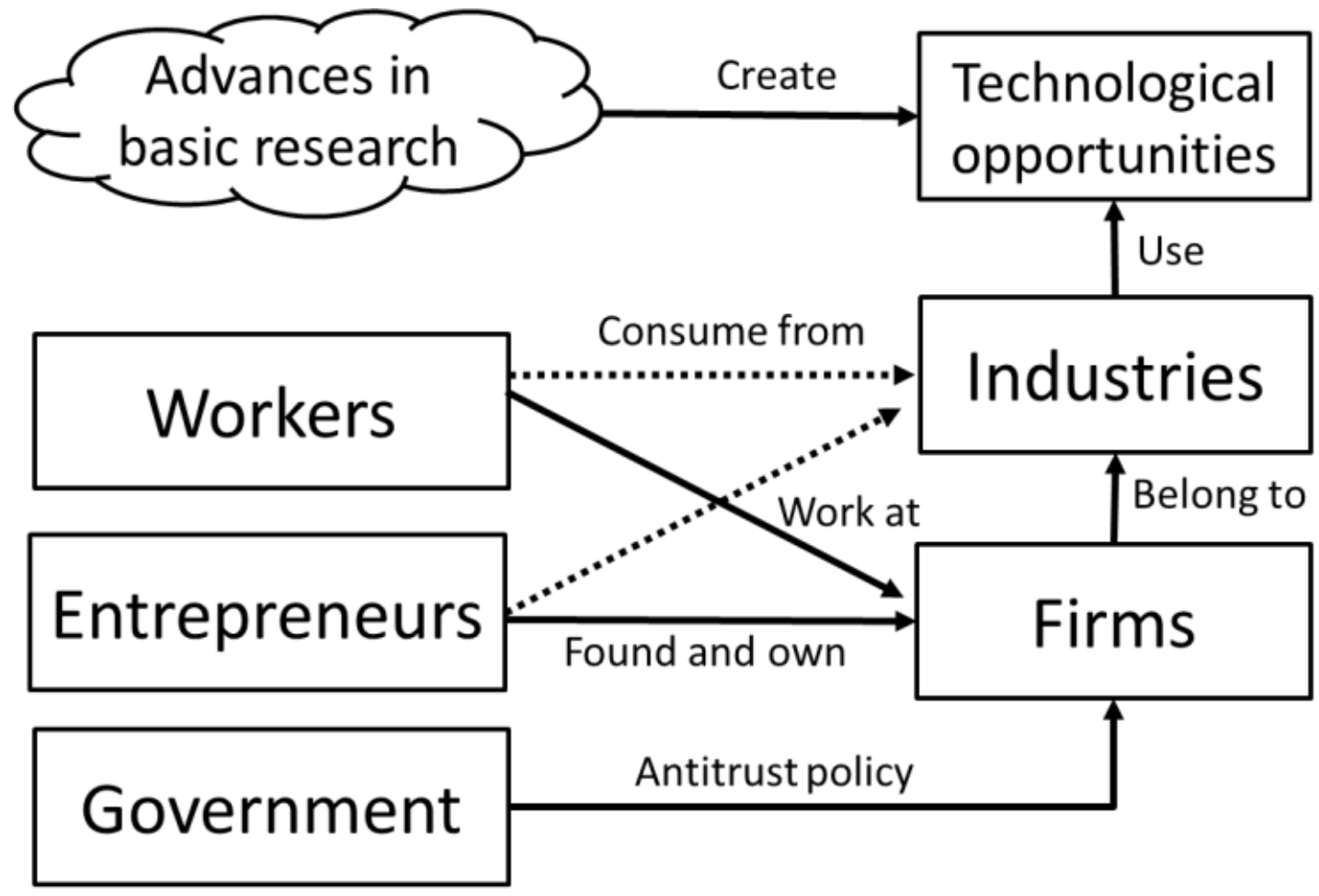

**Figure 1**: Overview of the model

My model is discrete-time. In each time step of the simulation:

1. Technological opportunities are updated.
2. Entrepreneurs decide whether they want to try to imitate or radically innovate.
3. Firms decide about their labor market demand.
4. Labor market interactions are processed.
5. Firms produce their output.
6. Firms perform research and development.
7. Firms pay wages and rents.
8. Households consume.
9. Firms calculate their profits.
10. Government may engage in antitrust activities.
11. Statistics are updated.

*2.1 Households and consumption*

The model is populated by $n_t$ households, which are split into two classes of agents who behave fundamentally differently: workers and entrepreneurs.

In the beginning of the simulation, $n_0^w$ agents are initialized as workers and $n_0^e$ agents are initialized as entrepreneurs. The number of entrepreneurs and workers in the model changes over time: On the one hand side, the number of workers grows every period by a parameter α. Entrepreneurs, on the other hand side, are in my model characterized by the fact that they own firms in a "1 to n" relationship, i.e., every entrepreneur may own multiple firms, but every firm is owned by a single entrepreneur. Hence, the ownership structure of firms in my model are more akin to "family businesses" than stock companies (in contrast to other work such as the model by Caiani et al. 2019). Due to this fact, I model an inheritance mechanism in which the number of entrepreneurs only grows if an entrepreneur who has at least two firms dies (which happens with the probability α). In this case, the firms (and savings) are split as evenly as possible between two new entrepreneurs. Finally, whenever the last firm of an entrepreneur goes bankrupt, she will become a worker. Hence, population growth is mostly exogenous, but the composition of the population in terms of the two classes is to a certain degree endogenous.

This model currently neither focuses on the workers and their behavior, nor on monetary quantities, which are therefore modeled in a highly simplified manner. Workers are employed by firms and receive a wage rate based on their employment. The wage rate that a worker would gain in a situation of full employment is set as the *numeraire*. New-born workers are endowed with one currency unit and I do not model debts or a financial sector, which ensures that the total number of currency units in the economy is equal to the total number of workers. This specification was chosen both because of its simplicity and because of its interesting implications: It retains a result from neoclassical microeconomics as a special case, namely that profits are only possible if there is some level of market concentration, which causes a lower-than-optimal level of output (and subsequently, unemployment). Hence, the existence of profits is per se inefficient from a static point of view. However, since profits allow entrepreneurs to save and found new firms, a possible trade-off between static efficiency and dynamic efficiency emerges.

Entrepreneurs own firms and obtain their profits. More specifically, firms pay out the difference between revenues in the last period and wage payments of the current period to their owner. [3]

I chose this specification to allow firms to use their profits to expand their production (only) if they believe that it is profitable to do so. Entrepreneurs save a fraction $\tau$ of their available funds (in order to found new firms at a later stage) and consume the rest. The funds of entrepreneur $y$ ( $f_{y,t}$) are thus updated according to the following equation, where $n_{y,t}^{i}$ denotes the number of firms owned by $y$, $p_{k,t-1}$ the price of the goods produced in this industry in the previous period, $q_{i,t-1}^{actual}$ the firm's output in the previous period, and $l_{i,t}$ the firm's labor force in the current period:

$$f_{y,t} = f_{y,t-1}\tau + \sum_{i}^{n_{y,t}^{i}} \left(q_{i,t-1}^{actual} p_{k,t-1} - l_{i,t}\right) \tag{1}$$

[3] If the economy was in a stationary equilibrium, this would be equal to the profits of the current period.

Since I abstract from upward social mobility and workers do not have any other incentive to invest, I assume – in line with much of the evolutionary agent-based literature (see, e.g., Dosi et al. 2010) – that workers consume all of their income. Hence, total nominal demand $d_t$ is given as follows, where $f_{y,t}$ denotes the funds of entrepreneur $y$, $n_t^w$ is the number of workers and $n_t^e$ is the number of entrepreneurs in the current period.

$$d_t = n_t^w(1 - u_t) + \sum_{y}^{n_t^e} (1 - \tau) f_{y,t} \tag{2}$$

Total nominal demand is then allocated to each active industry k (that observes a quantity which is larger than 0) according to its relative attractiveness. More specifically, industry-specific nominal demand $d_{k,t}$ is calculated by dividing its $\delta_k$ (which is drawn from a uniform distribution upon founding the industry) by the sum of the attractiveness values of all industries:

$$d_{k,t} = d_t \frac{\delta_k}{\sum_j \delta_j} \tag{3}$$

While this mechanism is simple, it is easily understandable, but still able to capture a salient feature of Schumpeterian thought, namely creative destruction: the foundation of a new industry tears away demand from the old industries by reallocating demand towards the new industry, thus causing a recession in old industries. Furthermore, it accounts for the facts that i) some industries produce goods that are just deemed more "important" by consumers and that ii) the relative importance of industries may be subject to change due to technological progress.

*2.2 Firms*

Firms hire workers to produce a specified good and to engage in research and development (R&D) activities. Each firm is confined to a single industry, but there may be more than one firm.

*2.2.1 Production planning*

If the number of firms operating in an industry is at least two, they engage in Cournot competition, a well-known model in which firms strategically choose their production quantity. In contrast to the standard model, however, firms are heterogeneous with respect to their labor productivity and do not possess perfect knowledge of the labor productivity of their competitors, as technological progress stochastically increases the labor productivity over time. Finally, firms also do not know the nominal demand faced by each industry in the upcoming period, since the emergence of new industries may decrease it (as discussed above). Firms hence have to form expectations about the behavior of their competitors and the evolution of demand.

As a heuristic, firms use OLS to try to predict the sum of the output of their competitors in the current period ($q_{i,t}^{comp}$) as well as the nominal demand that the industry faces in the current period ($d_{k,t}$), using an OLS regression that accounts for the exponential nature of the growth in output (due to technological change and population growth), as well as in demand (due to population growth), where $t°_t$ starts with 0 at the period in which this industry is born. They run a regression using the following model on a "memory" that consists of a list of all previous total output in the industry minus their own output:

$$\ln(y) = \ln(\beta_{y,0}) + \ln(\beta_{y,1}) t°_t + \varepsilon_{i,t} \tag{4}$$

Where y is either $q_{i,t}^{comp}$ or $d_{k,t}$, while $\beta_0$ is a constant, $\beta_1$ is the base of the exponential function, and $\varepsilon_{i,t}$ is an error term.

Predicted output and nominal demand in the period t is hence given by equation (5), where $\hat{y}$ is the prediction (i.e. $\widehat{q_{i,t}^{comp}}$ or $\widehat{d_{k,t}}$) and $\beta_{y,0}$ and $\beta_{y,1}$ are the respective regression estimates:

$$\hat{y} = \widehat{\beta_{y,0}} \widehat{\beta_{y,1}}^{t°_t} \tag{5}$$

After performing its predictions, each firm then computes the Cournot quantity $q_{i,t}^{Cournot}$ using its prediction and its labor productivity $a_{i,t}$ (the derivation of the Cournot quantity is shown in Appendix A).

$$q_{i,t}^{Cournot} = \sqrt{\widehat{d_{k,t}} \widehat{q_{i,t}^{comp}} a_{i,t}} - \widehat{q_{i,t}^{comp}} \tag{6}$$

This mechanism retains the (neo)classical general equilibrium result as a special case, namely in the absence of any technological progress. This is true, because firms will over time learn the true value of $d_{k,t}$, as well as the behavior of the competitors, which itself is the optimal response to their own behavior. This result is shown in Figure 2, which is to be found in the results section.

*2.2.2 Nominal labor demand*

Similar to Dosi et al. (2010), where firms invest a constant fraction of their revenue in R&D (based on an empirical stylized fact), firms in my model add a constant fraction $\nu$ of labor dedicated to research and development to the amount of labor necessary to produce the Cournot quantity:

$$l_{i,t}^{C} = \frac{q_{i,t}^{Cournot}}{a_{i,t}}(1+\nu) \tag{7}$$

Cournot competition fails to capture one crucial aspect of competition within my model: namely, the endogenous entry of new firms into an industry, which is incentivized by firms that produce low quantities (and thus reap a high rate of profit). Due to this fact, it is strategically rational for firms who command a large market share to produce a larger quantity than the Cournot quantity in order to deter possible entrants. This is also true for monopolists, where the Cournot quantity cannot be computed as it would be an infinitely small number.

This idea is captured by another simple heuristic: If the expected rate of profit from producing the Cournot quantity $r^{Cexpectation}$ lies above a rate of profit of $r^{target}$, the nominal labor demand is set to produce an expected rate of profit of $r^{target}$.

$$r^{Cexpectation} = \frac{q_{i,t}^{Cournot} \dfrac{\widehat{d_{k,t}}}{\left(q_{i,t}^{Cournot} + \widehat{q_{\iota,t}^{comp}}\right)} - l_{i,t}^{C}}{l_{i,t}^{C}} \tag{8}$$

$$l_{i,t}^{dmin} = \frac{q_{i,t-t}^{actual}\, p_{k,t-t}}{1 + r^{target}} \tag{9}$$

Actual labor demand $l_{i,t}^{d}$ is thus $l_{i,t}^{dmin}$ or $l_{i,t}^{C}$, whichever is higher:

$$l_{i,t}^{d} = \max\left(l_{i,t}^{dmin}, l_{i,t}^{C}\right) \tag{11}$$

Exceptions are made for firms which operate in a new-born industry. In the first time step of the simulation, nominal labor demand equals the entire funds a firm has at its disposal, i.e. the economy starts in a full-employment disequilibrium.

If a firm produces for the first time in a new-born industry after the first time step (i.e. after a radical innovation), it aims to achieve $r^{target}$ by producing $l_{i,t}^{dfounder}$ (see eq. 12), but is constrained by its funds $f_{i,t}$ (see eq. 13). I assume that a firm does not know the "attractiveness" of a new industry beforehand and thus possesses much less information about the demand for its product than firms in established industries. As a simple (and, on average, correct) heuristic, those firms assume that the new industry is as attractive as the average existing industry. Firms expect the total nominal demand observed over all industries of the previous period $d_{t-1}$ to grow by the population growth α. Firms expect that the number of industries grows by exactly one compared to the number in the previous period $n_{t-1}^{k}$ (i.e., that they are the only firm that successfully conducts a radical innovation in the respective period).

$$l_{i,t}^{dfounder} = \frac{\dfrac{d_{t-1}(1+\alpha)}{n_{t-1}^{k}+1}}{(1+r^{target})} \tag{12}$$

The actual number of workers employed by a firm i in period t is then given by its labor demand or its funds $f_{i,t}$, whichever is lower:

$$l_{i,t} = \begin{cases} \min\left(f_{i,t}, l_{i,t}^{dfounder}\right) & \text{for founders} \\ \min\left(f_{i,t}, l_{i,t}^{d}\right) & \text{for others} \end{cases} \tag{13}$$

A firm's funds, in turn, are given by its seed money (for founders) or by the revenue in the previous period (for others).

### *2.2.3 Research and development*

R&D activities in my model capture both innovation and imitation, and follow the standard logic as developed in the Keynes + Schumpeter model by Dosi et al. (2010) closely, with the exception of the imitation process, which I changed to account for the stylized fact that it seems to be harder to enter a technologically advanced industry than to enter an infant industry (as discussed in detail in the introduction). It is important to note that Dosi et al. (2010) also implemented a mechanism to incorporate “technological distance” between the imitating firm and the imitated ones. In their model, however, the technological distance does not affect the *overall* chance for a successful imitation, but the *target* of the imitation process, i.e., it is more probable to imitate a technologically close competitor. I, in contrast, assume that firms would always want to imitate their best competitor and that the chance to do so decreases with the technological distance between the two firms, as measured by the difference in labor productivities. In the appendix, I show that my results are not sensitive to this assumption, as the stylized facts are also replicated in a scenario where firms always try to imitate the firm which is closest to the productivity of itself (but still has a higher labor productivity than the respective firm).

Firms hire an additional fixed share $\nu$ of their production staff to engage in R&D (as reasoned in the previous subsection).

$$l_{i,t}^{RnD} = \frac{\nu l_{i,t}}{(1+\nu)} \tag{14}$$

Like in the evolutionary standard approach (Dosi et al. 2010), a fraction $\xi$ of the R&D staff is assigned to innovation, the rest to imitation.

$$l_{i,t}^{IN} = \xi l_{i,t}^{RnD} \tag{16}$$

$$l_{i,t}^{IM} = (1-\xi) l_{i,t}^{RnD} \tag{17}$$

Firms which are at the technological frontier dedicate all of their efforts to innovation, which is consistent with the empirics (Liao 2020).

The innovation process is given by a three-step process. First, a draw from a Bernoulli distribution decides whether a firm successfully invents a new production technology. The probability of a successful innovation, $\theta_{i,t}^{IN}$, is given as follows, where $l_{i,t}^{IN}$ denotes the number of workers who are assigned to innovation by firm i in period t and $\psi^{IN}$ is a parameter describing the innovation capability:

$$\theta_{i,t}^{IN} = 1 - e^{-\psi^{IN} l_{i,t}^{IN}} \tag{18}$$

From this equation, taken exactly from Dosi et al. (2010), it is clear that the probability of a successful innovation increases with the number of workers assigned to innovation, but that this increase is subject to diminishing returns.

The second step of the innovation process occurs if the Bernoulli draw was successful. The productivity change follows a multiplicative approach, where the rate of change $\Delta a_{i,t}^{*}$ is drawn from a normal distribution with the mean $\mu$ and the standard deviation $\sigma$. The productivity of a newly invented production technology $a_{i,t}^{*}$ is thus described by the following equation, where $a_{i,t-1}$ denotes the production technology used in the previous period:

$$a_{i,t}^{*} = a_{i,t-1}\left(1 + \Delta a_{i,t}^{*}\right) \tag{19}$$

In the third step, firms compare the invention with the existing technology. Only if $a_{i,t}^{*} > a_{i,t-1}$, i.e., $\Delta a_{i,t}^{*} > 0$, the invention will actually be adopted, i.e., cause an innovation, and become $a_{i,t}$, as only these inventions are economically meaningful by increasing labor productivity.

After processing innovation activities, firms try to imitate their best competitor. More specifically, they try to copy the most productive technology used in their industry during the last period. Similar to the innovation process, the probability of a successful imitation depends on a draw from a Bernoulli distribution. However, in contrast to the innovation mechanism and the previous literature (e.g. Dosi et al. 2010), the probability for a successful imitation decreases if the imitating firm and the imitation target are technologically more distant.

The reason for this choice that as it seems to be harder to copy technologies that are far more advanced and it seems to be more difficult to enter a technologically advanced industry. This assumption is motivated by anecdotal evidence from the IT industry such as the market for operating systems, where it would be extremely difficult to enter, e.g., the market for mobile operating systems successfully beside Apple and Android. [4] However, this assumption is also supported by empirical evidence, as Bloom et al. (2013) show that "technology spillovers" (i.e. via imitation) are more likely between firms which are located closely within the technology space.

[4] Obviously, this fact does not only stem from technological reasons, i.e., it is not only because it takes a lot of (sunk) costs to create another operating system that is on par with the incumbents, but also from, e.g., network effects (think of app compatibility etc.) and user habits. While my mechanism does not explicitly account for these reasons, it can very well serve as a proxy for them.

I formalize this idea as follows: The probability for a successful imitation is given by $\theta_{i,t}^{IM}$, which depends on number of workers who are assigned to imitation ($l_{i,t}^{IM}$), an imitation capability parameter, $\psi^{IM}$ as well as the firm's labor productivity after computing the three-step innovation process described in the preceding paragraphs $a_{i,t}$, the highest labor productivity in industry k (i.e. the industry in which firm i is active) in the previous period $\bar{a}_{k,t-1}$ and an imitation distance penalty parameter $\vartheta$.

$$\theta_{i,t}^{IM} = 1 - e^{\frac{-\psi^{IM}}{\left(1+\vartheta\left(\bar{a}_{k,t-1}-a_{i,t}\right)\right)} l_{i,t}^{IM}} \tag{20}$$

If the imitation is successful and more productive than the currently used technology (even after accounting for a possible invention), the firm adopts this technology.

*2.2.4 Production*

After accounting for R&D, those workers who are not assigned to R&D, i.e., $l_{i,t}^{prod}$, produce consumption goods using the technology described by the labor productivity $a_{i,t}$.

$$q_{i,t}^{actual} = l_{i,t}^{prod} a_{i,t} \tag{21}$$

*2.3 Labor market*

The labor market is highly simplified as it is not the focus of my current analysis. The nominal labor demand of firms has already been described in subsection 2.2.2. As already mentioned, the nominal wage rate of one unit of labor (i.e., the labor provided by a worker at full employment in one period) serves as the *numeraire*. I assume that workers can work without restrictions in any industry. Hence, the (average) nominal wage rate per worker is given by the employment rate.

*2.4 Technological opportunities*

Technological opportunities represent a potential product space, which is exogenously expanded by advances in basic research and which can endogenously be exploited by entrepreneurs to found new industries. This assumption is in line with Schumpeterian theory (see Freeman 1982). Technological opportunities are given by a $x \times y$ matrix, where each element of the matrix represents one potential industry. Scientific advances may change $x$ and $y$. More specifically, $x$ may increase with the probability $\varphi^x$ and $y$ with the probability $\varphi^y$ at the beginning of each time step of the simulation. This assumption implies that technological progress comes in waves that usually become larger, as this assumption allows the product space to grow superlinearly. A maximum size for $x$ and/or $y$ can be set in order to rigorously explore the impact of the arrival of new technological opportunities or to confine the waves to a certain maximum size. If, for instance, $x$ is limited to 1, scientific advances always only add a single new technological opportunity. A technological opportunity may become an actual industry, if an entrepreneur successfully carries out a radical innovation, which is described in the next subsection.

### *2.5 Entrepreneurs*

Entrepreneurs are at the heart of the "Schumpeter Mark I" engine of this model. During each time step, they try to found a new firm by either a) imitating in an industry where they do not yet own any firm (if such an industry exists) or by b) radically innovating by founding a new industry in an unexploited technological opportunity (if such a technological opportunity exists). The probability of a successful radical innovation is given by $\rho$.

In deciding what to do, entrepreneurs compare the expected profit of their investment, which is equal to their current funds $f_{y,t}$. The expected profit of pursuing a radical innovation $E\left(\Pi_{y,t}^{radical}\right)$ is given by the probability to successfully conduct a radical innovation $\rho$, the expected wage payments, i.e., the expected nominal labor demand of the firm $l_{i,t}^{d}$ and the expected nominal demand of the industry $d_{k,t}^{expected}$:

$$E\left(\Pi_{y,t}^{radical}\right) = \rho(d_{k,t}^{expected} - l_{i,t}^{d}) \tag{22}$$

Entrepreneurs do not know a priori how successful a new market will be. Analogous to eq. (12) and in line with the model mechanics, they hence expect it to be on average as important as any other sector. Expected demand is hence given by the following equation, where $d_{t-1}$ denotes total nominal demand of the past period, α is the population growth and $n_{t-1}^{k}$ the number of industries with an output which is larger than 0:

$$d_{k,t}^{expected} = \frac{d_{t-1}(1+\alpha)}{n_{t-1}^{k}+1} \tag{23}$$

The nominal labor demand of the new firm founded by a radical innovation is given by the following equation, where $r^{target}$ denotes the target (analogous to eq. 12 and 13):

$$l_{i,t}^{d\ radical} = \min\left(f_{y,t}, \frac{d_{k,t}^{expected}}{1 + r^{target}}\right) \tag{24}$$

The expected value of pursuing an imitation is more demanding, as we have to calculate the imitation probability, as well as the Cournot quantity. The imitation probability follows the imitation mechanism at the firm-level, but replaces $l_{i,t}^{IM}$ with 1 (i.e., the entrepreneur) and the imitation capability parameter with an own $\psi^{ENT}$ that takes into account that entrepreneurs may have a particularly high ability to conduct R&D:

$$\theta_{y,t}^{IM} = 1 - e^{\frac{-\psi^{ENT}}{\left(1+\vartheta\left(\bar{a}_{k,t-1}-a_0\right)\right)}} \tag{25}$$

The expected value of trying to imitate $E\left(\Pi_{y,t}^{imitation}\right)$ is thus given by the following equation, as firms expect the nominal demand to stay constant:

$$E\left(\Pi_{y,t}^{imitation}\right) = \theta_{y,t}^{IM}\left(d_{k,t-1} - l_{i,t}^{d}\right) \tag{26}$$

Labor demand is given by equation (27):

$$l_{i,t}^{d} = \min\left(f_{y,t}, l_{i,t}^{C}\right) \tag{27}$$

where $l_{i,t}^{C}$ is calculated in the same way as it is for existing firms (see eq. 4-7).

Entrepreneurs will only become active once they are able to attract at least one worker with their savings, i.e., they need to accumulate at least some savings to be able to found a new sector. Entrepreneurs' saving behavior was already described in subsection 2.1 (eq. 1).

### *2.6 Government*

In this simple model, the government only serves a single function, namely that it can engage in antitrust activities. If a firm i's market share exceeds a policy variable $m_t^{max}$, it will be broken up into two separate companies with a probability of $1 - \omega^{o_{i,t}}$, where $o_{i,t}$ is the number of periods that the firm has exceeded the maximum market share $m_t^{max}$ (which is reset to 0 as soon as the firm is below $m_t^{max}$) and ω is a parameter governing the likelihood of antitrust action. Technically, a new firm j will be founded that copies the production technology of i and receives half of its funds. Furthermore, the firm's "memory" which is the basis for the OLS learning of the expected output of competitors is reset to the previous period, where it is the total output minus half of the output of firm i for the "new" and the "old" firm.

## 3 Results

In subsection 3.1, I first illustrate how the model "works" using by analyzing single simulation runs that show how i) the OLS learning mechanism relates to the Nash equilibrium of the Cournot oligopoly, and ii) the impact of the "Schumpeter Mark I" vs. the "Schumpeter Mark II" engines on output and inequality.

I then calibrate the unified model to a baseline scenario using approximate Bayesian computation (subsection 3.2). I show in subsection 3.3 that the calibrated model reasonably captures important indicators of the empirics of growth and inequality in the US quantitatively and show that the baseline calibration is able to replicate all ten stylized facts regarding "declining business dynamism" since the 1980s proposed by Akcigit and Ates (2021).

Finally, I explore the impact of three counterfactual scenarios to explore how declining business dynamism is driven by technological factors vs. antitrust policies. My analysis shows that stronger antitrust enforcement is most effective at combatting "declining business dynamism" and is able to reverse at least seven out of 10 stylized facts. Technological factors, on the other hand, play a much more limited role in my model.

The model is implemented in NetLogo (Wilensky 1999) and available as open source. [5] I analyze the model outputs with the ggplot2 (Wickham 2016) package for R (R Core Team 2022).

### *3.1.1 OLS learning and Nash equilibrium*

This subsection is dedicated to illustrating how the model works. To do so, I begin in a hypothetical situation in which neither the Schumpeter Mark I nor Schumpeter Mark II engine are active and the economy only has a single industry. Furthermore, I set the population growth to zero. Figure 2 shows the evolution of output over time when starting from a situation of disequilibrium compared to a situation in which the Nash equilibrium is set from the start (the mathematical derivation of the Nash equilibrium can be found in Appendix A).

[5] https://github.com/patrickmellacher/schumpetermark1plus2

This result shows that, in absence of any growth and innovation, the OLS learning heuristic pursued by the firms approximates the Nash equilibrium of the Cournot model in the long term, although the adaption process takes considerable time if the number of firms is large. This exercise shows that the standard equilibrium result is indeed a *special case* of my more general model.

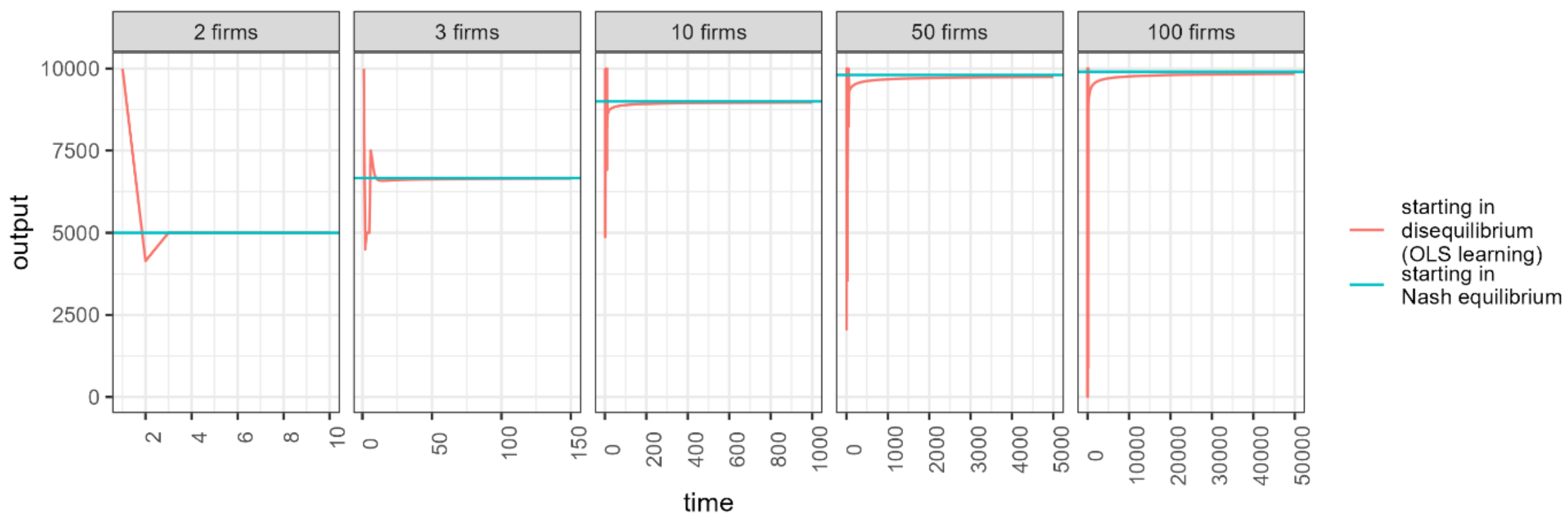

Figure 2: OLS learning vs. Nash equilibrium (note that the x-axis differs for each plot!)

### *3.1.2 Schumpeter Mark I only*

We now activate the Schumpeter Mark I engine by allowing entrepreneurs to save a fraction of their funds (in this case 0.2) and found new firms – either by radically innovating (i.e., by creating a new industry) or by imitating an existing firm (i.e., entering a specific industry). The simulation is shown in Figure 3 and starts with only a single active industry (called industry 8). Over time, entrepreneurs accumulate funds from their profits gained in this industry, which allows them to found new industries. This process in turn reduces the economy-wide importance of industry 8 by “tearing away” its demand (i.e., creative destruction).

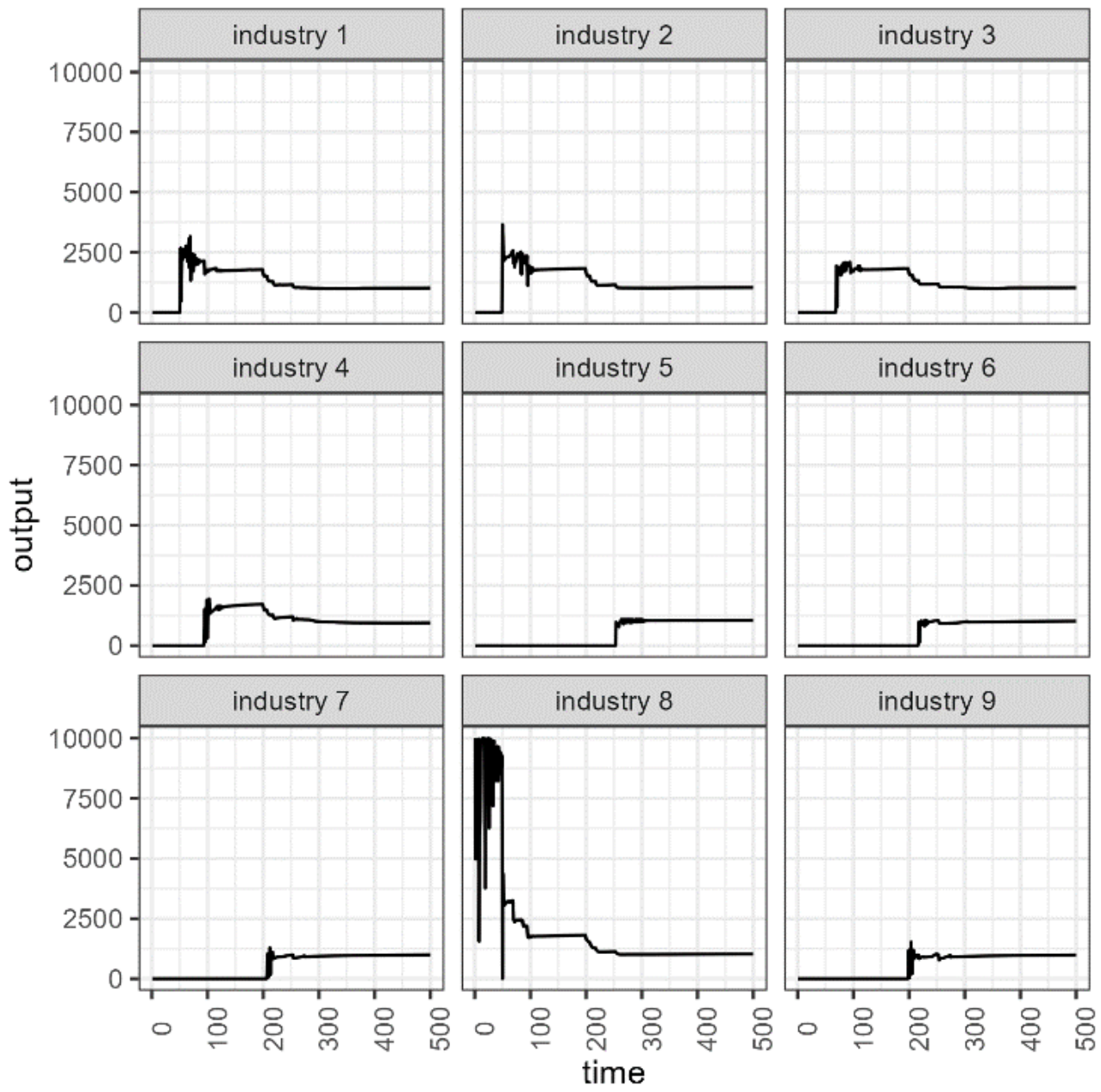

Figure 3: An economy with 9 (potential) industries over time where only the Schumpeter Mark I engine is active

### *3.1.3 Schumpeter Mark II only*

We then “turn off” the Schumpeter Mark I engine again by setting the entrepreneurial savings rate to 0, but activate the Schumpeter Mark II engine by setting the R&D propensity to 2%. The illustrative runs in Figure 4 show that this mechanism creates exponential growth, but also allows for periods of recession. Recessions are triggered by the fact that the growth expectations of firms are not met exactly due to the stochastic nature of the innovation process. Figure 4 furthermore suggests that growth is higher if the number of firms in a given industry is higher (and the market concentration is hence less concentrated). This is due to the fact that the number of workers employed in R&D in my model increases with the number of workers employed in the production process.

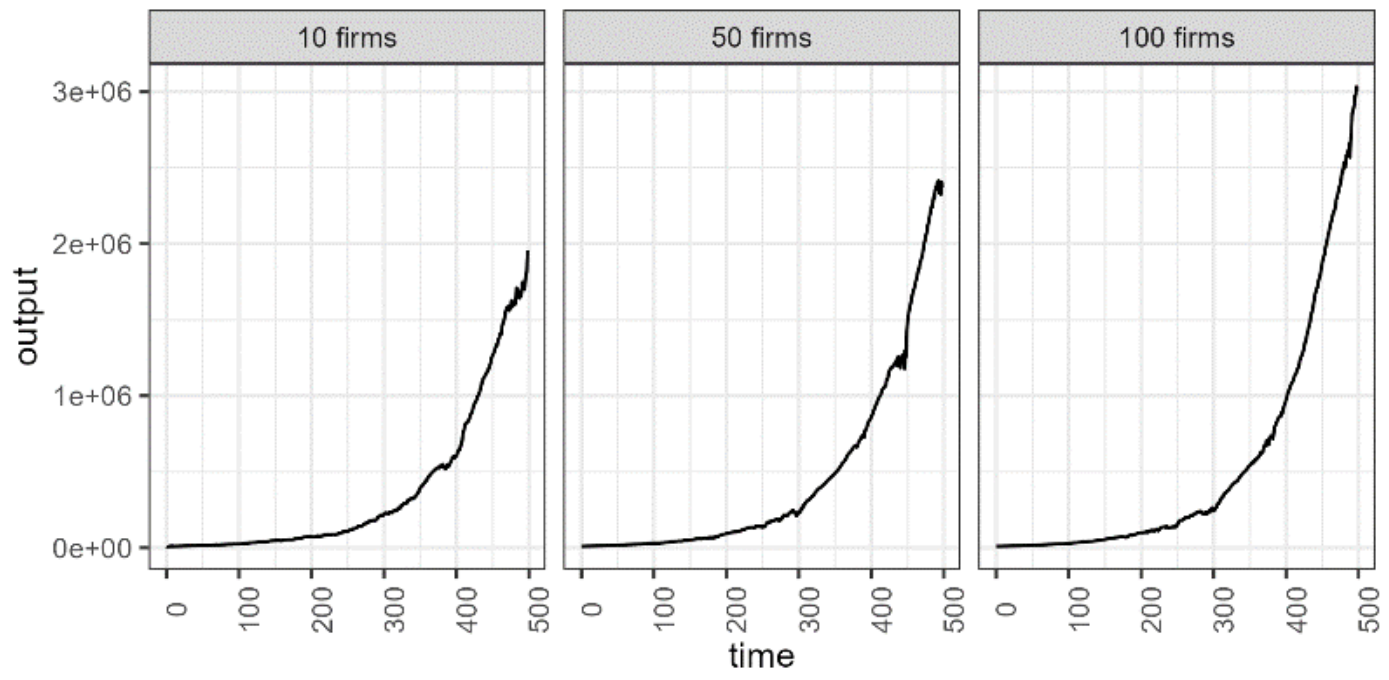

Figure 4: An economy with a single industry over time with three different initial firm numbers.

### *3.1.4 Schumpeter Mark I + II unified*

Finally, I activate both the Mark I and the Mark II "engines" by allowing entrepreneurs to save funds (in order to found new firms) and by allowing firms to invest in R&D. Figure 5 illustrates output, market concentration and the consumption share for a fixed number of nine (potential) industries, starting with industry 1. All parameters (except for the fixed number of potential industries and a lower number of entrepreneurs of 50) are set according to the empirical calibration described in the next subsection. We can see that the model allows for a great deal of heterogeneity between industries, as highly concentrated industries can co-exist with moderately concentrated ones. Likewise, some industries observe a very high growth rate in output, while others are unimportant with regard to the whole economy. Both growth and market concentration are influenced by the "attractiveness" of the product produced by the specific industry, which determines the consumption share for each industry. If households spend a large fraction of their income on a specific industry, these revenues allow firms to hire more workers for R&D (thus increasing the growth rate), but also attract new competitors (thus pushing down market concentration, which also increases growth, as discussed in subsection 3.1.3).

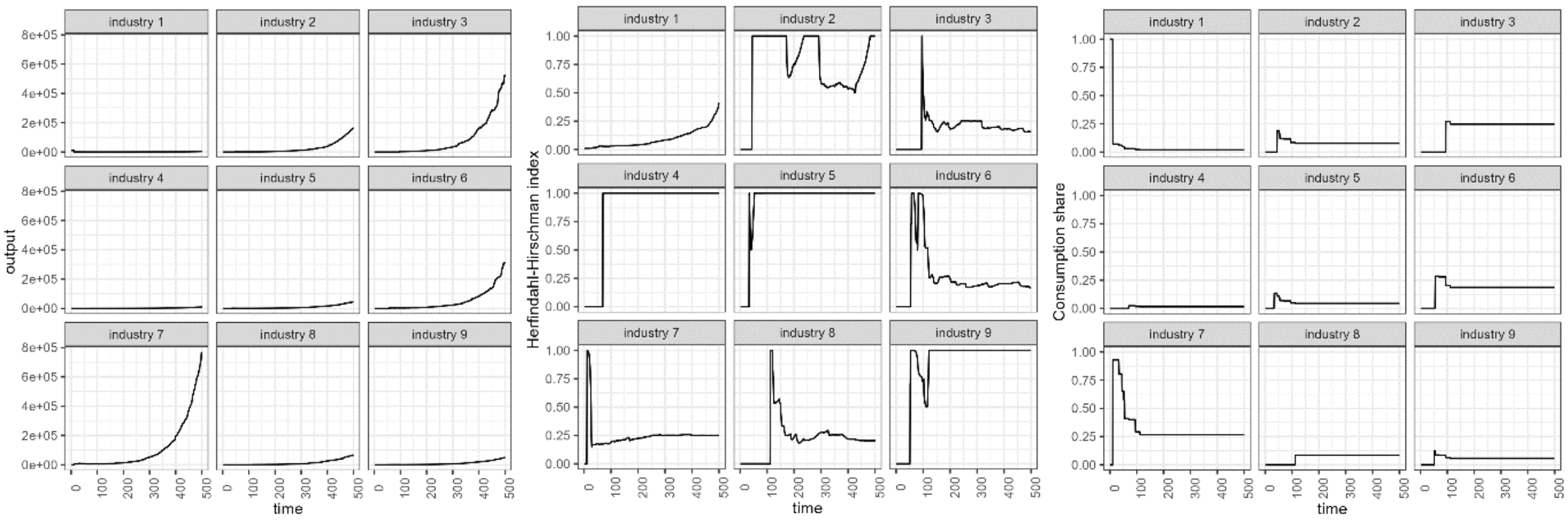

Figure 5: Output (left), market concentration (center) and consumption share (right) in the Unified Schumpeter Mark I + II model with a fixed number of 9 (potential) industries.

### *3.2 Quantitative calibration*

After analyzing the most important components of the model separately, I turn to the analysis of multiple runs (i.e., Monte Carlo simulations, see Dosi et al. 2010). In order to do so, I calibrate the model to quantitatively match the growth rate of the US economy from 1820 until 2018 (from the Maddison Project Database, Bolt and Van Zanden 2020) and the levels of inequality observed in 1980 and 2018 measured as the pre-tax national income received by the top 1% of the US population (from the World Inequality Database, Chancel et al. 2022).

In order to calculate a growth rate, one faces the problem that the economy faces multiple (incommensurable) consumption goods and that the composition of the consumer basket changes over time (as shown in subsection 3.1.4). In order to tackle this problem, I approximate the growth rate of GDP per capita in the model by using the average productivity growth rate (which is weighted according to the market share of each firm $m_{i,t}$) per industry and the total employment growth rate:

$$g_t = \frac{l^s_{x,t}}{l^s_{x,t-1}} \frac{\sum_i m_{i,t} \frac{a_{i,t}}{a_{i,t-1}} - 1}{n^k_t} \tag{28}$$

In empirically calibrating the model, I use the "freelunch" package for R developed and presented by Carrella (2021). In order to do so, I first set up a baseline policy scenario which focuses on i) antitrust which arguably witnessed a "golden age" going ca. from the 1940s until the 1970s (see Kovacic and Shapiro 2000) as ii) the period of "declining business dynamism" and rising inequality from the 1980s onwards. The "golden age" of antitrust in my model is characterized – in a stylized way – by the fact that firms are not allowed to exceed a certain market share. Following the Chicago school's argument developed in the 1970s (and assumed to become effective in 1980) that market power alone is insufficient to justify antitrust action, this policy is inactive from 1980 onwards.

Every period in the model represents one quarter in the real world. The model starts with 300 periods as a burn-in phase, after which the model starts in a virtual year 1820. Table 1 shows the timeline of events in the baseline scenario.

**Table 1: Baseline scenario**

| Period | Event |
|---|---|
| 300 | End of the burn-in phase (corresponding to the year 1820). |
| 780 | Start of the "market power" approach to antitrust. Firms may be split up into two separate companies if they exceed a certain market share (corresponding to the year 1940). |
| 940 | End of the "market power" approach. Firms may no longer be split up due to their market share (corresponding to the year 1980). |
| 1108 | End of the simulation (corresponding to the year 2018). |

After setting up the baseline scenario, I fit the model's parameters to match the empirically observed growth rate (from 1820 until 2018) and income inequality (in 1939, 1980 and 2018) in the USA. To do so, I run the model 30,000 times where four parameters are drawn from a uniform distribution: i) the standard deviation of an invention, ii) the target profit rate, iii) the imitation distance penalty parameter and iv) the maximum market share during the "golden age of antitrust". I used all eight algorithms (such as random forest and various forms of approximate Bayesian computation (ABC)) to find the algorithm which is best-equipped to fit the model outputs to the empirical data. A cross validation exercise shows that the neural network ABC approach proposed by Blum and Francois (2010) achieves the best average performance in a 5-fold cross validation built into the "freelunch" package by Carrella (2021).

The population growth parameter is directly calibrated from the empirical data (Bolt and Van Zanden 2020) as the mean annual population growth of the USA between 1820 and 2018. Other parameters were chosen for convenience (making the number of entrepreneurs and industries not "too large" as this increases the computational cost of running the model).

I then simulate the model based on the estimates from the neural network ABC approach. Table 1 shows the results, Table 2 shows the calibrated parameter values. It is notable that the estimated maximum market share (ca. 15%) seems to be very low. However, in practice market shares which were below this bound were considered to be problematic at times: for instance, in the 1960s a merger that would have resulted in a market share of 5% was invalidated by the courts (Sawyer 2019).

**Table 2: Calibrated parameter values**

| Parameter | Symbol | Value | Comment |
|---|---|---|---|
| initial number of workers | $n_0^w$ | 10000 | - |
| initial number of entrepreneurs | $n_0^e$ | 100 | 1 % of workers |
| likelihood of a radical innovation | $\rho$ | 0.01 | - |
| propensity to invest in R&D | $\nu$ | 0.02 | - |
| population growth | $\alpha$ | 0.004418605 | Annualized from Bolt and Van Zanden (2020) |
| imitation capability parameter | $\psi^{IM}$ | 0.3 | - |
| invention capability parameter | $\psi^{IN}$ | 0.3 | - |
| imitation capability parameter (entrepreneurs) | $\psi^{ENT}$ | 3 | - |
| antitrust parameter | $\omega$ | 0.8 | - |
| share of R&D staff assigned to invention | $\xi$ | 0.5 | - |
| initial technological opportunities occupied | - | 1 | - |
| initial labor productivity | $a_0$ | 1 | - |

| propensity of entrepreneurs to save | $\tau$ | 0.2 | - |
|---|---|---|---|
| initial technological opportunities (x) | $x$ | 2 | - |
| initial technological opportunities (y) | $y$ | 2 | - |
| probability for x to grow | $\varphi^x$ | 0.008 | - |
| probability for y to grow | $\varphi^y$ | 0.008 | - |
| invention mean | $\mu$ | 0 | - |
| invention standard deviation | $\sigma$ | 0.0114918 | fitted |
| imitation distance penalty | $\vartheta$ | 150.3954 | fitted |
| target profit rate | $r^{target}$ | 0.3321438 | fitted |
| maximum market share when antitrust is active | $m_t^{max}$ | 0.149987 | fitted |

### *3.3 Empirical Validation*

This section is devoted to evaluating whether. In the first subsection, I evaluate the output of 1,000 simulation runs using the parameter calibration fitted as described above with the empirical counterparts (i.e., "descriptive output validation" as discussed by Tesfatsion 2017).

#### *3.3.1 Quantitative results*

Table 3 shows the results of the descriptive output validation. While the calibrated model very slightly overestimates average GDP growth and slightly overestimates the income inequality in 1939, income inequality in 1980 and 2018 are located in the confidence interval given by the simulated mean +/- the standard deviation. While it may be possible to further improve the model's fit in a computationally expensive way, it is also very well possible that the model overestimates the level of income inequality in 1939 because the model does not account for other important policies and dynamics that may have affected the level of inequality, such as the "new deal" policies enacted in the 1930s of the US.

**Table 3: Empirical vs. simulated values (baseline scenario)**

| **Variable** | **Empirical value (Source)** | **Simulated mean (standard deviation)** |
|---|---|---|
| Average growth rate GDP 1820-2018 | 0.01632015 (Bolt and Van Zanden 2020) | 0.01770003 (0.001276319) |
| Income share of entrepreneurs (income share of the top 1 %) 1939 | 0.1957 (Chancel et al. 2022) | 0.218457 (0.01698936) |
| Income share of entrepreneurs (income share of the top 1 %) 1980 | 0.1043 (Chancel et al. 2022) | 0.1293587 (0.04674446) |
| Income share of entrepreneurs (income share of the top 1 %) 2018 | 0.1925 (Chancel et al. 2022) | 0.1900389 (0.02734596) |

### *3.3.2 Empirical stylized facts*

In this subsection, I explore the model's ability to reproduce salient empirical stylized facts on the industry, as well as on the macro level.

*Industry-level Stylized Facts (SF):*

This model is primarily centered around the birth and evolution of new industries. Hence, its ability to replicate industry-level stylized facts is important. Although not all industries evolve equally, a characteristic pattern of industry evolution can be found in manufacturing industries, which is characterized by three industry-level stylized facts (**SFI**) that are robustly reproduced in the model (see fig.):

1.) The number of firms over time follows an inverted U-curve pattern (**SFI1**): it first increases rapidly, before declining again, after which it becomes constant (Gort and Klepper 1982, Klepper and Graddy 1990).
2.) The price decreases (**SFI2**, Klepper and Graddy 1990).
3.) The output increases (**SFI3**, Klepper and Graddy 1990).

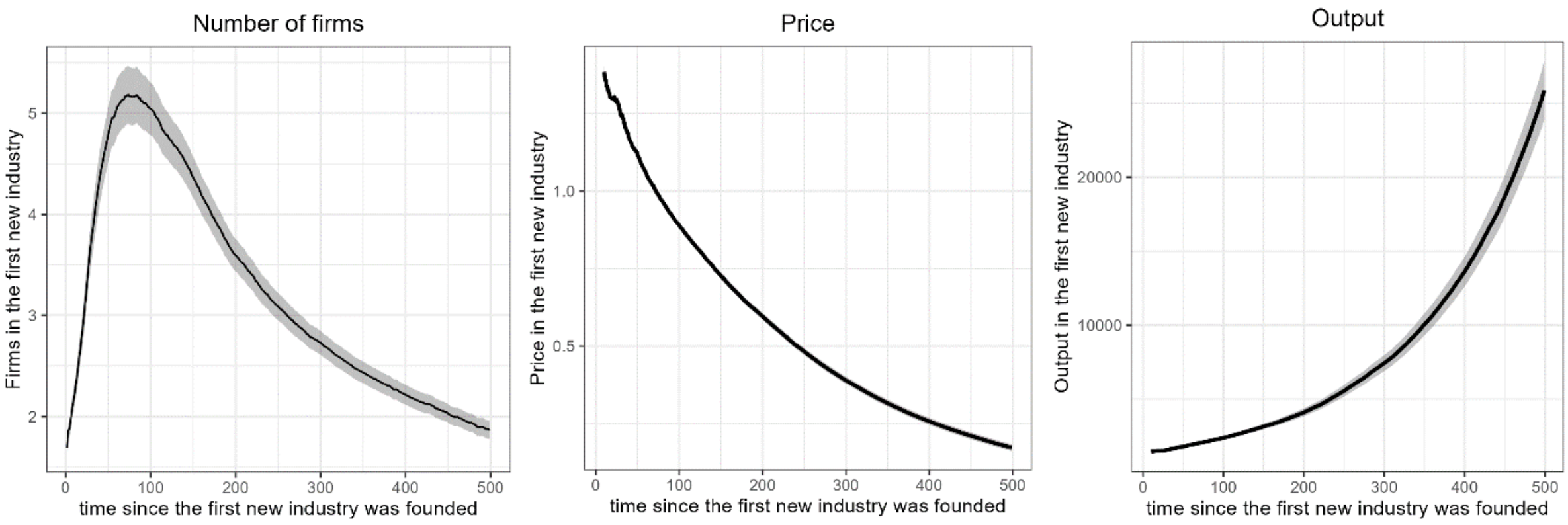

Figure 6: number of firms (left), the price (center) and output (right) for the first new industry founded in the baseline scenario (mean value and 95% confidence interval)

However, there are also two industry-level stylized facts that the model cannot robustly reproduce (i.e., it only reproduces them within certain time frames), namely that this the decrease in price and the increase in output follow a falling rate (see Klepper and Graddy 1990 and Dinlersoz and MacDonald 2009). Whether output rises (and, conversely, the price falls), is driven by two factors: technological progress and competition.

On the one hand side, higher levels of market concentration achieved during the “shakeout phase” in which the number of firms decreases imply that a lower number of workers are assigned to R&D, which can translate to a lower level of productivity growth, which in turn affects output and prices. On the other hand, this effect is counteracted by the fact that output decreases due to a higher level of market concentration in the “shake out” phase that starts relatively soon. In order to robustly account for these stylized facts, the model could be changed such that technological progress in mature industries becomes weaker (or less likely), either due to technological reasons (e.g., because exploiting further technological opportunities may become more difficult, see Saviotti and Pyka 2004a) or due to economic reasons (e.g., because oligopolists and, in particular, monopolists may have a weaker incentive to invest in R&D).

*Macro-level stylized facts: Growth*

The model features two macro-level stylized facts (**SFG**) on growth which are highlighted by Saviotti et al. (2020):

1.) The productive efficiency in each sector increases over time (**SFG1**).
2.) The number of distinct active industries increases over time (**SFG2**).

Both stylized facts were already shown to be true in Figure 5). The model currently does not account for a third stylized fact highlighted by Saviotti et al. (2020), namely that the quality of products produced in each sector increases over time. While changing the R&D process to account for this stylized fact would be straightforward, the resulting necessary changes in the demand function (and, subsequently, in the expectation formation) seem to be much more complicated to realize within the currently used Cournot framework.

*Macro-level stylized facts: Inequality*

The model features two stylized facts on inequality highlighting the prevalence, dynamics and persistence of inequality:

1.) The level of inequality in society first increased, then decreased (Kuznets 1955) after increasing again in the 1980s (e.g., Piketty 2014, **SFIN1**).
2.) There is persistent inequality not only between social classes but also within the group of the "rich", i.e., in this model the entrepreneurs / firm owners. While the exact distribution is disputed, e.g., whether it follows a Pareto law (see Chan et al. 2017), it is indisputable that there is significant heterogeneity and inequality within the top 1%, and that the top 0.1% of the population have played a very important role in rising inequality (see, e.g., Saez and Zucman 2016), **SFIN2**).

The two stylized facts are illustrated in

**Figure 1**Figure 7, which shows the funds held by entrepreneurs over time, as well as the wealth distribution within the class of entrepreneurs in the final simulation step.

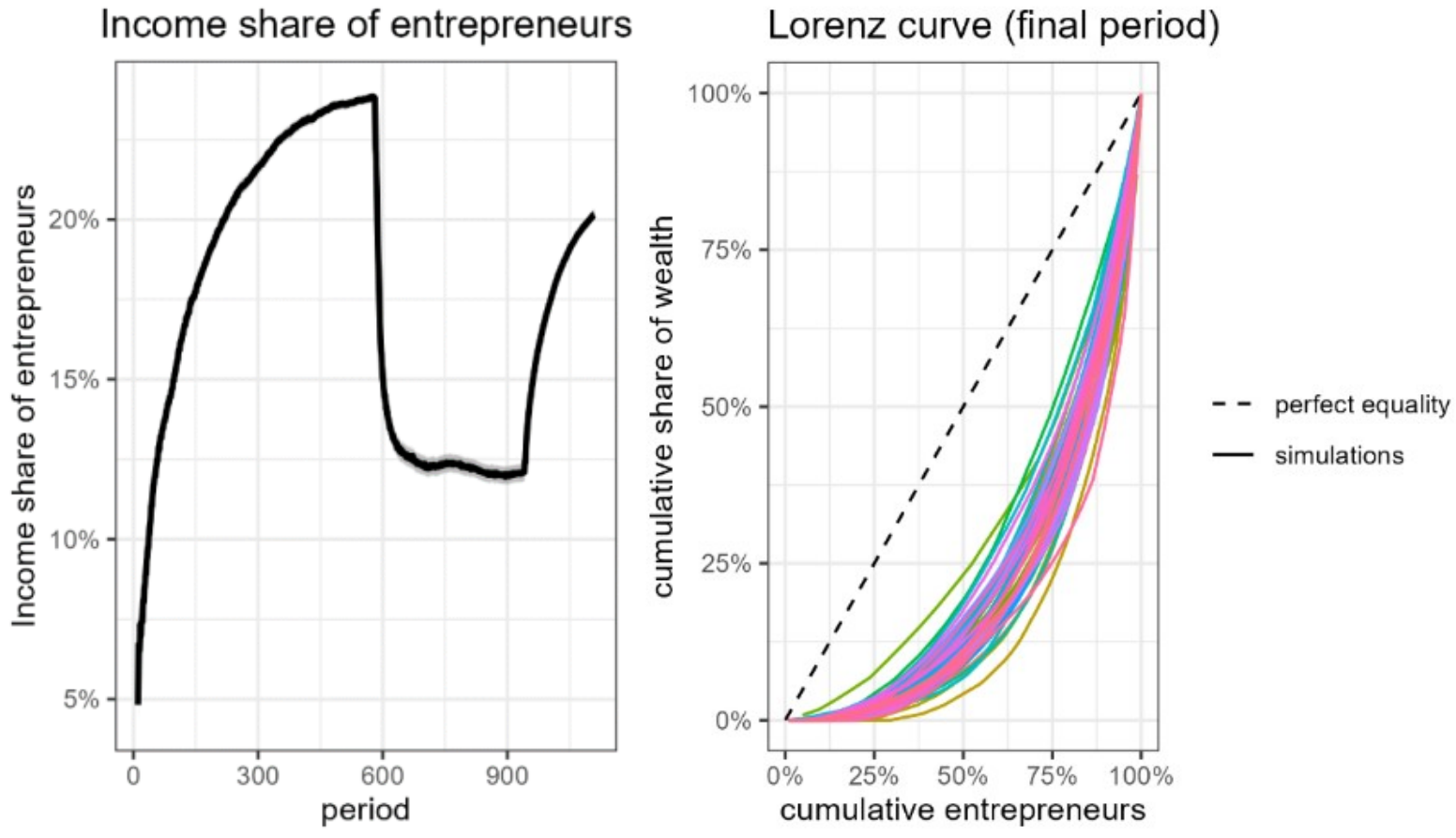

Figure 7: Share of income held by entrepreneurs (mean and 95% CI starting at period 10 for 1,000 simulation runs) and Lorenz curves (perfect equality vs. actual distribution in 50 simulation runs) in the final period for the baseline calibration.

*Macro-level stylized facts: Declining business dynamism*

Finally, this model is able to replicate ten stylized facts on "declining business dynamism" stated by Akcigit and Ates 2021 (except for the text in the parantheses, the following list is a direct quotation from their paper):

1.) Market concentration has risen (**SFDBD1**, measured in the model by the average Herfindahl-Hirschman Index).
2.) Average markups have increased (**SFDBD2**, measured in the model by the sum of profits divided by the sum of costs).
3.) Average profits have increased (**SFDBD3**, measured in the model by the sum of profits divided by GDP).
4.) The labor share of output has gone down (**SFDBD4**, measured in the model by the sum of wages divided by GDP).
5.) The rise in market concentration and the fall in labor share are positively associated (**SFDBD5**, shown as a linear model and a Pearson correlation coefficient between the average Herfindahl-Hirschman Index and the wage share measured by the sum of wages divided by GDP).
6.) The labor productivity gap between frontier and laggard firms has widened (**SFDBD6**, measured in the model by one minus the mean productivity laggard firms relative to the productivity of industry leaders as mean across all industries).
7.) Firm entry rate has declined (**SFDBD7**, measured in the model by the number of new firms in a given period).

8.) The share of young firms in economic activity has declined (**SFDBD8**, measured in the model by the number of workers employed by firms which are younger than 20 periods = 5 years divided by the total number of employed workers).

9.) Job reallocation has slowed down (**SFDBD9**, measured in the model by the sum of job creation and destruction divided by the sum of employed workers).

10.) The dispersion of firm growth has decreased (**SFDBD10**, measured by the standard deviation of employment growth of firms).

Figure 8 shows how all of the ten stylized facts are replicated in the baseline scenario using data from the simulated year 1980 (i.e., period 940) onwards:

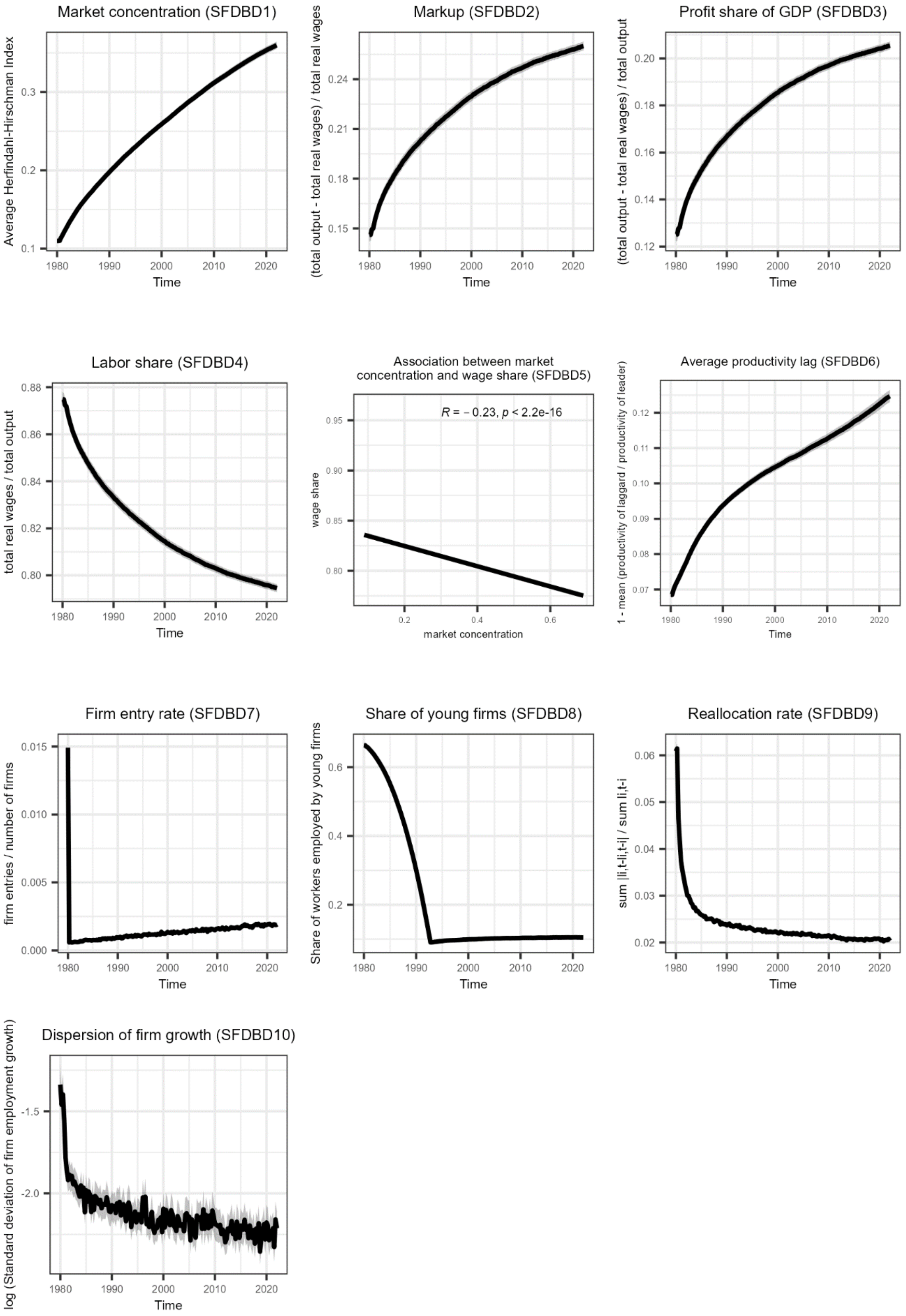

Figure 8: Replication of all of the ten stylized facts on declining business dynamism since the 1980s in the calibrated baseline scenario (means + 95% Confidence Intervals for 1,000 simulation runs).

### *3.4 Counterfactual experiments: Explaining declining business dynamism*

After showing that the empirically calibrated baseline scenario replicates all of the ten stylized facts on "declining business dynamism" (Akcigit and Ates 2021), I conduct counterfactual policy simulations in order to study the drivers of declining business dynamism. In particular, I am interested in whether and how the decline in business dynamism can be overcome in specific dimensions in my model. In order to do so, I study three counterfactual scenarios and compare them against the baseline scenario for each stylized fact:

*Longer antitrust:*

In this scenario, the "market power" approach remains active until the end of the simulation. Otherwise, this scenario behaves exactly like the baseline scenario.

*More technological opportunities:*

In this scenario, the probability that the potential product space grows in each dimension ($\varphi^x$, $\varphi^y$) is increased from 0.8% to 1.2% per period. Otherwise, this scenario behaves exactly like the baseline scenario.

*No technological distance penalty:*

In this scenario, there is no technological distance penalty (starting from the initial period). Otherwise, this scenario behaves exactly like the baseline scenario.

Table 4 summarizes key macroeconomic indicators for the baseline scenario vs. the counterfactual scenarios. Like in the general equilibrium model by Vaziri (2022), antitrust policies foster economic growth. In contrast to her results, however, antitrust policy does not represent a quasi-Pareto improvement where workers are better off and firm owners are unaffected. Instead, entrepreneurs / firm owners face lower real profits in this scenario.

Since my model captures the endogenous entry of firms as well as an endogenous markup and endogenous technological change, my results suggest that there is no trade-off between dynamic and static efficiency with regard to antitrust as hypothesized by authors such as Sidak and Teece (2009), who explicitly refer to neo-Schumpeterian theories and evolutionary economics. Instead, there is a conflict of interest between firm owners and workers regarding antitrust.

The scenario that allows for more technological opportunities witnesses slightly lower productivity growth and inequality, which translates to a statistically significant (but rather small) decrease in real profits (as measured by the logarithm of the total sum of goods produced), and a significant increase in the number of industries.

The scenario that does not exhibit a technological distance penalty witnesses the lowest inequality and the highest productivity growth rate. However, this growth is limited to a much smaller number of industries. This is due to the fact that it is often more profitable for entrepreneurs in this scenario to enter existing industries than to establish new ones.

**Table 4: Stylized facts in the counterfactual scenarios**

| Indicator | Senarios | | | |
|---|---|---|---|---|
| | **Baseline Scenario** | **Longer antitrust** | **More technological opportunities** | **No technological distance penalty** |
| productivity growth since 1980 (quarterly) | 0.00543551 (0.0006934138) | 0.00592447 *** (0.0005624012) | 0.005394271 *** (0.0006629517) | 0.0111527 *** (0.001024534) |
| Income share of entrepreneurs (final period) | 0.1947444 (0.02661249) | 0.1269425 *** (0.04457196) | 0.1867047 *** (0.02320483) | 0.05580841 *** (0.02043319) |
| Number of industries (final period) | 112.1045 (43.07363) | 109.716 (40.74547) | 119.937 *** (42.7129) | 36.87564 *** (5.905981) |
| Log total output (final period, sum over all industries) | 17.44818 (0.4508959) | 17.63384 *** (0.4676885) | 17.43291 (0.444134) | 23.33622 *** (0.5544082) |
| Log real wages (final period) | 17.22619 (0.4709231) | 17.49208 *** (0.4905428) | 17.221 (0.4621155) | 23.28361 *** (0.5607039) |
| Log real profits (final period) | 15.82231 (0.3812664) | 15.53886 *** (0.4440587) | 15.76759 *** (0.376365) | 20.30037 *** (0.5484762) |

Note: *** deviation from the baseline scenario $p<0.001$(Welch's test)

The next natural step is to investigate whether the stylized facts regarding an increase in inequality (Piketty 2014) and a decline in business dynamism (Akcigit and Ates 2021) since the 1980s are replicated in the model. Table 5 summarizes the results, which are shown in detail in Figure 9.

The "longer antitrust" scenario, in which strict antitrust enforcement remains in place from the 1940s throughout the simulation, clearly reverses seven out of ten stylized facts, with another stylized fact (namely that the firm entry rate has declined) remaining inconclusive. Only two stylized facts, namely that the share of young firms in economic activity has declined and that the reallocation rate has gone down, are still replicated in this scenario, albeit at lower levels than witnessed in the baseline scenario. These results suggest that antitrust policy is indeed highly effective at combatting "declining business dynamism".

Technological factors, on the other hand, are much less powerful in combatting the facts on "declining business dynamism" in my model. The scenario that allows for a stronger growth of technological opportunities (i.e., allows for a larger potential product space) still replicates all of the ten facts, although some effects are dampened. The scenario in which the firms' imitation probability does not decrease with the imitation distance, i.e., where no imitation distance penalty exists, is able to reverse three stylized facts on declining business dynamism (as well as the stylized fact of an increase in inequality). Generally, however, the *levels* in this scenario differ drastically from the other scenarios (as well as the real world) throughout the simulation runs and should thus not be taken as realistic estimates, but rather to deliver an impression of expected qualitative changes.

**Table 5: Stylized facts in the counterfactual scenarios**

| Stylized Fact | Description | Senarios | | | |
|---|---|---|---|---|---|
| | | Baseline Scenario | Longer antitrust | More technological opportunities | No technological distance penalty |
| **SFIN1** | Inequality increased since the 1980s | ✓ | - | ✓ | - |
| **SFDBD1** | Market concentration has risen. | ✓ | - | ✓ | ✓ |
| **SFDBD2** | Average markups have increased. | ✓ | - | ✓ | - |
| **SFDBD3** | Average profits have increased. | ✓ | - | ✓ | - |
| **SFDBD4** | The labor share of output has gone down. | ✓ | - | ✓ | - |
| **SFDBD5** | The rise in market concentration and the fall in labor share are positively associated. | ✓ | - | ✓ | ✓ |
| **SFDBD6** | The labor productivity gap between frontier and laggard firms has widened. | ✓ | - | ✓ | ✓ |
| **SFDBD7** | Firm entry rate has declined. | ✓ | ~ | ✓ | ✓ |
| **SFDBD8** | The share of young firms in economic activity has declined. | ✓ | ✓ | ✓ | ✓ |
| **SFDBD9** | Job reallocation has slowed down. | ✓ | ✓ | ✓ | ✓ |
| **SFDBD10** | The dispersion of firm growth has decreased. | ✓ | - | ✓ | ✓ |

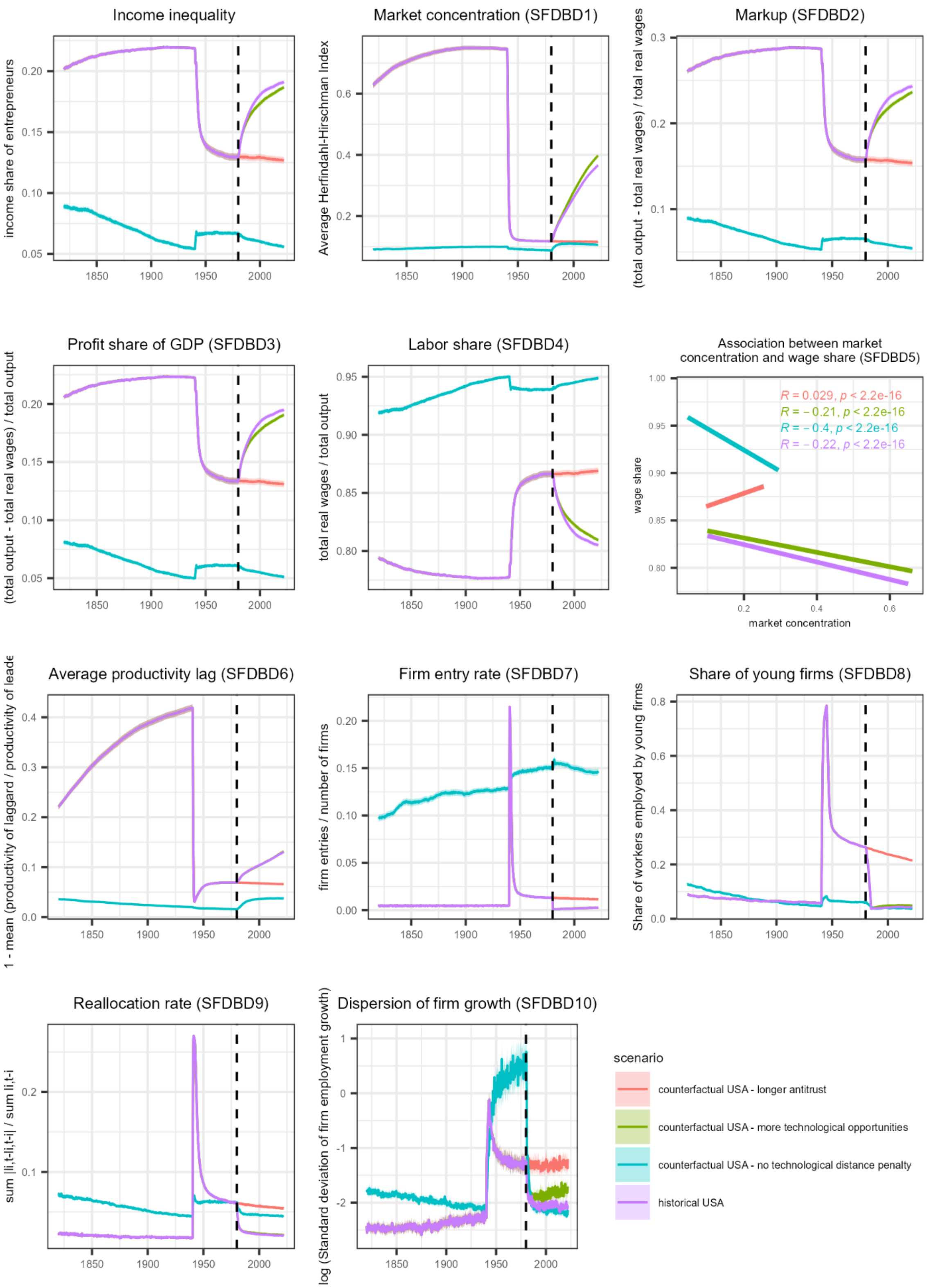

**Figure 9**: Replication of stylized facts (mean + 95% Confidence Interval) in the various counterfactual scenarios

## 4 Conclusion

In this paper, I developed a simple agent-based model that incorporates key features of the so-called Schumpeter Mark I and the Schumpeter Mark II models as termed by Freeman (1982). My unified Schumpeter Mark I + II model allows for endogenous technological change in two dimensions: i) the number of (consumption good) industries, which can be increased by actions taken by explicitly modeled entrepreneurs (i.e., the "Schumpeter Mark I" engine), and ii) the firm-specific labor productivity, which is affected by R&D processes conducted in firms (i.e., "Schumpeter Mark II"). In doing so, the model incorporates different notions of technological change, such as "incremental innovation" (in this model on the firm-level) vs. "radical innovation" (on the level of creation of new industries coming in clusters, see Perez 2007), as well as "product innovation" (through creating new industries producing new products) vs. "process innovation" (through producing goods in a more efficient way).

Using a reference table algorithm (see Carrella 2021), I calibrated the model to reasonably reproduce key features of the US economy regarding growth and inequality quantitatively. I further showed that, despite its simplicity, the calibrated model is able to reproduce a large number of stylized facts regarding i) the industry life-cycle, ii) growth, iii) inequality, as well as iv) "declining business dynamism" since the 1980s (Akcigit and Ates 2021) in its baseline calibration.

I then conducted three counterfactual policy simulations to study whether increasing inequality and "declining business dynamism" since the 1980s was reproduced in my model due to i) technological reasons, focusing on the arrival of new technological opportunities the ease of imitation, or ii) political reasons, focusing on the role of antitrust, or iii) reasons that cannot be captured by such parameters and hence represent the "nature" of capitalism within my model or the model's limits.

I found that a scenario where antitrust policy is as stringent as it was during its "golden age" from the 1940s to the 1970s (i.e., the "Chicago school" of antitrust never gains any power) is most effective at combatting "declining business dynamism and reverses seven out of ten stylized facts mentioned by Akcigit and Ates (2021) and at least alleviates the three other stylized facts.

In contrast, scenarios that are more conducive to technological change are only able to reduce the worst impacts of declining business dynamism. While the scenario which allows for more radical innovations replicates all stylized facts (albeit sometimes to a quantitatively lower degree), eliminating the "imitation distance penalty" parameter, which impedes the ability of firms and entrepreneurs to imitate firms which are technologically more advanced, reverses declining business dynamism in only three out of ten dimensions.

At least two stylized facts are reproduced in any scenario, namely that the job reallocation rate decreases (i.e., the sum of job creation and destruction over the total number of jobs) and that the economic share of young firms decreases (measured by the number of workers employed in firms which are younger than five years old). In my model, these tendencies are generally always present except for a short time after implementing antitrust policies. One possible modification of the model to circumvent these general tendencies would be to allow for a demise of whole industries. While established industries in my model experience a relative decline in importance when new industries are created, they will not vanish completely, as it sometimes happens in the real world, e.g., in the case of videocassette recorders.

Other limitations of this model are i) a missing finance sector (see, e.g., Assenza et al. 2015 for a prominent agent-based account of this topic), ii) a highly stylized labor market (see, e.g., Dawid et al. 2009 for a much more complex approach), iii) a stylized goods market and consumption rules (see, e.g., Dosi et al. 2022 for a recent sophisticated evolutionary approach), iv) the absence of any government policies other than antitrust such as patents (see e.g. Dosi et al. 2023), or labor market reforms (see, e.g., Dawid et al. 2019) or taxes. It would hence be interesting to study the relationship between antitrust, but in particular also other government policies with the stylized facts on declining business dynamism in a more holistic model.

**Acknowledgements**: A previous version of this manuscript was included as "Growth, Concentration and Inequality in a Unified Schumpeter Mark I + II model" as part of my PhD thesis. I am thankful for the helpful comments by participants of seminar presentations that I gave at Sant'Anna and the Catholic University of Milan (both in 2022), the conference of the International Schumpeter Society 2021, as well as the conference of the European Society for the History of Economic Thought 2021, the conference of the European Association for Evolutionary Political Economy 2020, the workshop Agent-based economics at the University of Graz in 2019 (especially Herbert Dawid, Andrea Roventini and Claudius Gräbner), and the doctoral school of economics Graz. I thank Andrea Roventini, Christian Gehrke and David Haas for comments and suggestions for improvement on previous versions of this manuscript. Any errors are mine.

**Appendix A: Derivation of the Cournot quantity and equilibrium**

This section derives the Cournot condition (used in eq. 6 in the main text) as well as the equilibrium quantity under the condition of constant and uniform productivity (used in subsection 3.1.1 of the results section)

Firms want to maximize profit Π is given by revenue R minus costs C:

$$\Pi = R - C \tag{A1}$$

Revenue is given by the price multiplied by the quantity. Price, in turn, depends on the total nominal demand for this industry $d_{k,t}$ which is divided among the total output of the firms producing in this industry (i.e., own output $q_{i,t}$ plus output of the competitors $q_{i,t}^{comp}$) Costs are given by workers employed in the production process multiplied by their wage rate. The wage rate is the numeraire (=1), while the workers employed in the production process are given by the own output divided by the labor productivity $a_{i,t}$. Hence, eq. (A1) transforms into eq. (A2):

$$\Pi = q_{i,t} \frac{d_{k,t}}{q_{i,t} + q_{i,t}^{comp}} - \frac{q_{i,t}}{a_{i,t}} \tag{A2}$$

Maximization of profits requires the first derivative (see eq. A3) to be 0 (see eq. A4)

$$\frac{\delta \Pi}{\delta q_{i,t}} = \frac{d_{k,t}(q_{i,t} + q_{i,t}^{comp}) - d_{k,t} q_{i,t}}{\left(q_{i,t} + q_{i,t}^{comp}\right)^2} - \frac{1}{a_{i,t}} \tag{A3}$$

$$a_{i,t} d_{k,t} (q_{i,t} + q_{i,t}^{comp}) - a_{i,t} d_{k,t} q_{i,t} - \left(q_{i,t} + q_{i,t}^{comp}\right)^2 = 0 \tag{A4}$$

Rewriting (A4) leads to (A5):

$$q_{i,t}{}^2 + 2 q_{i,t} q_{i,t}^{comp} + q_{i,t}^{comp2} + a_{i,t} d_{k,t} q_{i,t}^{comp} = 0 \tag{A5}$$

Using the reduced quadratic equation leads to (A6):

$$q_{i,t_{1,2}} = -\frac{2 q_{i,t}^{comp}}{2} \pm \sqrt{\left(\frac{2 q_{i,t}^{comp}}{2}\right)^2 - \left(q_{i,t}^{comp2} + a_{i,t} d_{k,t} q_{i,t}^{comp}\right)} \tag{A6}$$

Since $q_{i,t} > 0$, $q_{i,t}^{comp} > 0$, $a_{i,t} > 0$, $d_{k,t} > 0$, we can rewrite (A6) to (A7):

$$q_{i,t} = -q_{i,t}^{comp} + \sqrt{\left(a_{i,t} d_{k,t} q_{i,t}^{comp}\right)} \tag{A7}$$

Eq. (A7) directly leads to eq. (6) in the main text.

In order to obtain the equilibrium solution for this Cournot game under the condition that demand stays constant and labor productivity is constant and uniform across $n_k$ competitors, we have to rewrite $q_{i,t}^{comp}$ as $n_k q_{i,t}$, which leads to eq. (A8):

$$q_{i,t} = -n_k q_{i,t} + \sqrt{\left(a_{i,t} d_{k,t} n_k q_{i,t}\right)} \qquad (A8)$$

Rewriting and squaring yields (A10):

$$((n_k + 1) q_{i,t})^2 = a_{i,t} d_{k,t} n_k q_{i,t} \qquad (A9)$$

Dividing by $q_{i,t}$ and rewriting leads to (A11)

$$q_{i,t} = \frac{a_{i,t} d_{k,t} n_k}{(n_k + 1)^2} \qquad (A10)$$

Since demand for the goods produced in industry k is given by total demand (which is equal to the number of workers) multiplied with the expenditure share for industry k (i.e., the attractiveness value for k divided by the sum of all attractiveness values of the economy), we can substitute $d_{k,t}$ and find (A12), which is used as a comparison baseline in subsection 3.1.1 of the results section):

$$q_{i,t} = \frac{a_{i,t} n_0^w \frac{\delta_k}{\sum_j \delta_j} n_k}{(n_k + 1)^2} \qquad (A11)$$

## Appendix B: Robustness check imitation target

Figure A 1 **Fehler! Ungültiger Eigenverweis auf Textmarke.**shows that all ten stylized facts on "declining business dynamism" are also replicated if we do not assume that firms always want to imitate the competitor with the highest labor productivity, but the "next" competitor in the productivity ranking.

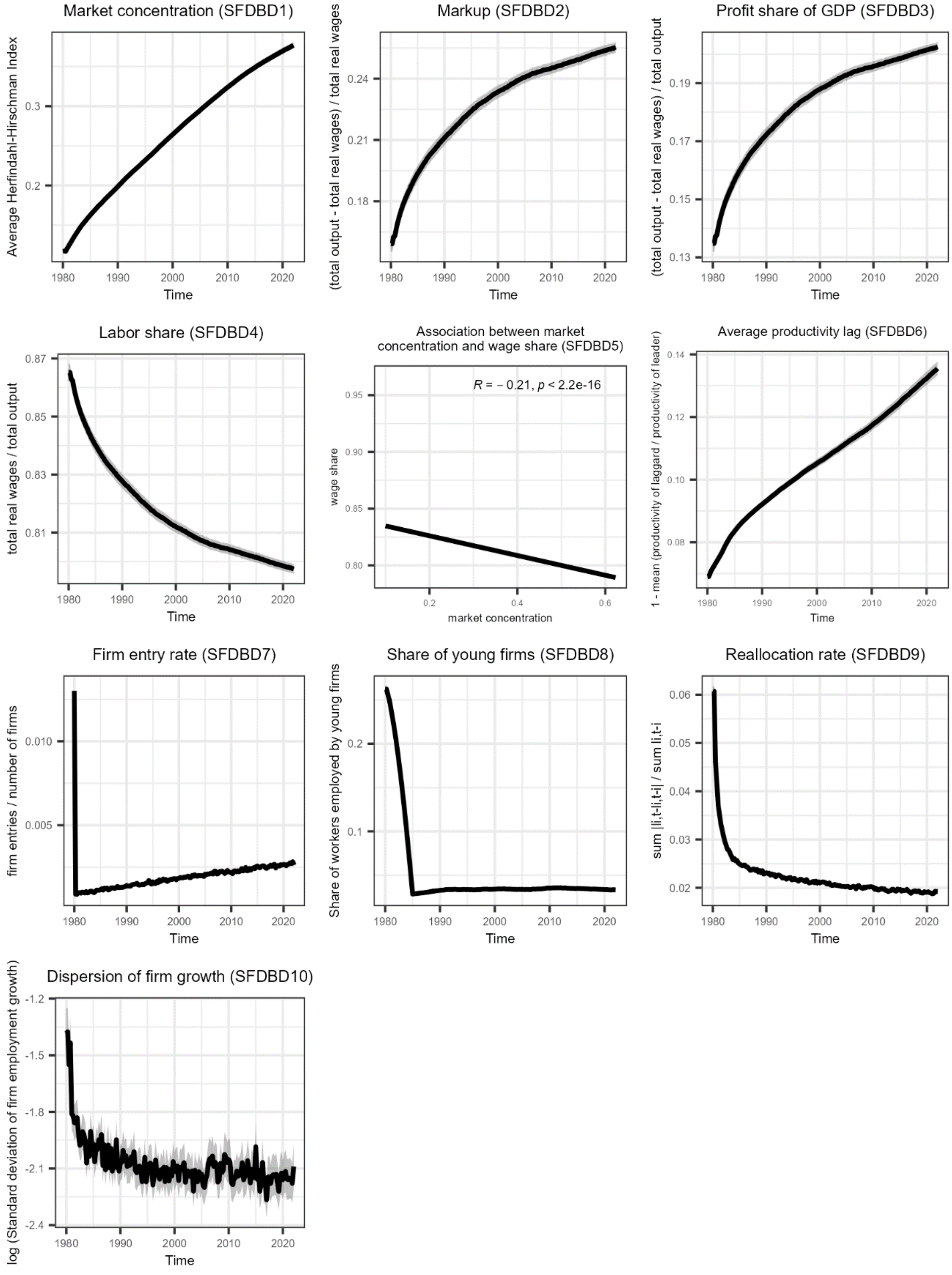

Figure A 1: Stylized facts on „declining business dynamism" with a different imitation target.